\documentclass[a4paper,twocolumn,11pt]{quantumarticle}
\pdfoutput=1

\makeatletter
\def\@ptsize{1}  % 1=11pt to match your document class option
\makeatother

\usepackage[utf8]{inputenc}
\usepackage[english]{babel}
\usepackage[T1]{fontenc}
\usepackage{amsmath,amssymb}
\usepackage{hyperref}
\usepackage{float}
\usepackage{tikz}
\usepackage{pgfplots}
\pgfplotsset{compat=1.18}
\usepgfplotslibrary{groupplots}
\usepackage{booktabs} 
\usetikzlibrary{shapes.geometric, arrows.meta, positioning,fit,patterns}
\usepackage{placeins} 

\usepackage{algorithm} 

\usepackage{algpseudocode}
\usepackage{etoolbox}

\usepackage{caption}
\usepackage{pgfplots} 
\pgfplotsset{compat=1.18} 
\usetikzlibrary{shapes, arrows.meta, positioning}
\usepackage{titlesec}
\usepackage{listings}
\usepackage{xcolor}
\usepackage{tabularx}
\usepackage{array}
\usepackage{ragged2e}
\usepackage{fontawesome}
\usepackage{enumitem}
\usepackage{graphicx}
\usepackage{multirow}
\usepackage{tikz}
\usetikzlibrary{arrows.meta}  % Add to preamble
\usetikzlibrary{shapes.geometric, backgrounds, fit, positioning}

\definecolor{codegreen}{rgb}{0,0.6,0}
\definecolor{codegray}{rgb}{0.5,0.5,0.5}
\definecolor{codepurple}{rgb}{0.58,0,0.82}
\definecolor{backcolour}{rgb}{0.95,0.95,0.92}

\lstdefinestyle{mystyle}{
    backgroundcolor=\color{backcolour},   
    commentstyle=\color{codegreen},
    keywordstyle=\color{magenta},
    numberstyle=\tiny\color{codegray},
    stringstyle=\color{codepurple},
    basicstyle=\ttfamily\footnotesize,
    breakatwhitespace=false,         
    breaklines=true,                 
    captionpos=b,                    
    keepspaces=true,                 
    numbers=left,                    
    numbersep=5pt,                  
    showspaces=false,                
    showstringspaces=false,
    showtabs=false,                  
    tabsize=2
}

\lstdefinestyle{style1}{
    basicstyle=\ttfamily\small,
    backgroundcolor=\color{white},  
    breaklines=true,
    frame=single,
    linewidth=\textwidth
}

\makeatletter
\renewcommand\subparagraph{%
   \@startsection{subparagraph}{5}{\parindent}%
   {3.25ex \@plus1ex \@minus .2ex}%
   {-1em}%
   {\normalfont\normalsize\bfseries}%
}
\makeatother
\begin{document}

\title{A fidelity-driven approach to quantum circuit partitioning via weighted hypergraphs for noise-resilient computation}

\author{Awad Wehbe}
%\orcid{0000-0002-2445-2701}
\email{81920001@students.liu.edu.lb}
\author{Safiya Al Khatib}
%\orcid{0000-0002-2445-2702}
\email{22030262@students.liu.edu.lb }
\author{Abdel-Mehsen Ahmad}
%\orcid{0000-0003-2796-4806}
\email{abdelmehsen.ahmad@liu.edu.lb}
\affiliation{School of Engineering, Lebanese International University, Bekaa, Lebanon}
\begin{abstract}
Noisy Intermediate-Scale Quantum (NISQ) devices, hampered by high error rates (e.g., CNOT $\epsilon_{\text{CNOT}} = 0.05$) and limited connectivity, demand effective circuit partitioning. Standard heuristics often overlook gate-specific error impacts, yielding suboptimal divisions with excessive communication overhead and degraded fidelity. We introduce \texttt{Fidelipart}, a novel framework that recasts quantum circuits as fidelity-aware hypergraphs. In this model, gate error rates and structural dependencies dictate node (gate) and hyperedge (multi-qubit gate and qubit-line temporal connectivity) weights, compelling an Mt-KaHyPar partitioner to minimize cuts through error-prone operations.
Evaluating \texttt{Fidelipart} against BQSKit's \texttt{QuickPartitioner} on benchmarks (up to 24-qubit/88-gate) under a linear topology with consistent local re-mapping, we demonstrate marked superiority. \texttt{Fidelipart} achieved 77.3\%–100\% reductions in SWAP gates (estimated logical realignments) and up to 52.2\% fewer cut qubits. These physical improvements yielded substantial estimated fidelity gains: for a 6-qubit circuit, fidelity surged by over 243.2\% (from 0.1724 to 0.5916 with $\epsilon_{\text{CNOT}}=0.05$). Similar significant SWAP reductions in larger circuits promise corresponding fidelity enhancements. While \texttt{Fidelipart} incurred a modest 8-13\% runtime increase and variable depth impact, its profound fidelity improvements underscore the critical advantage of integrating fine-grained error-awareness for robust NISQ computations. Source code: \url{https://github.com/awadwehbe/fidelipart}.

\end{abstract}

\section{Introduction}
\label{sec:introduction}
Quantum computing holds the potential to revolutionize fields such as cryptography, optimization, and materials science by harnessing quantum mechanical principles like superposition and entanglement \cite{nielsen2010}. However, executing quantum circuits on Noisy Intermediate-Scale Quantum (NISQ) devices, with limited qubits (50--100), high gate error rates (e.g., CNOT error rates, $\epsilon_{\text{CNOT}}$, assumed to be around 0.05 for weighting purposes in our model), and constrained connectivity, requires sophisticated compilation techniques \cite{preskill2018, nam2020}. Circuit partitioning, a critical step in quantum compilation, decomposes complex circuits into subcircuits to manage resource constraints, mitigate errors, and enable parallel execution \cite{murali2019, li2019}.

Effective partitioning must minimize cross-partition dependencies (cut qubits), reduce communication overhead (e.g., SWAP gates for logical realignment), and preserve circuit fidelity, particularly for highly entangled circuits where errors accumulate rapidly \cite{cowtan2019}. Existing methods, such as BQSKit's \texttt{QuickPartitioner} \cite{bqskit2021} (when processed with a consistent physical mapping strategy), can incur significant SWAP gate overhead (e.g., 44 SWAPs for our 24-qubit, 88-gate ``Circuit L'' benchmark) and consequently suffer from reduced fidelity (e.g., 0.1241 for ``Circuit L'' based on an evaluation $\epsilon_{\text{CNOT}}=0.05$), with partitioning and re-mapping times for large circuits in the range of seconds (e.g., 4.159s for ``Circuit L'').

We introduce \texttt{Fidelipart}, a fidelity-aware hypergraph-based partitioning framework integrated into BQSKit, designed to address these challenges. \texttt{Fidelipart} models quantum circuits as hypergraphs, with nodes representing gates. Hyperedges, weighted by factors derived from gate error rates (e.g., giving higher penalty to cuts involving CNOTs with $\epsilon_{\text{CNOT}} = 0.05$ compared to single-qubit gates like Hadamards with $\epsilon_H = 0.001$), capture multi-qubit gate connectivity and temporal dependencies along qubit operational sequences \cite{schlag2021}. It employs a workflow involving: (1) hypergraph conversion, (2) partitioning using Mt-KaHyPar\cite{Heuer2021mtkahypar} to create balanced subcircuits, (3) optional partition merging based on shared qubit thresholds to minimize cut qubits, and (4) dependency graph construction to ensure correct execution order. Benchmarked on circuits ranging from 6 to 24 qubits, \texttt{Fidelipart} achieves SWAP gate reductions (estimated logical realignments) of up to 100\% (e.g., 0 vs. 8 for our 6-qubit, 22-gate ``Circuit S'', and 10 vs. 44 for ``Circuit L''), and up to 52.2\% fewer cut qubits (e.g., 11 vs. 23 for ``Circuit L''). This leads to substantial fidelity improvements, with estimated fidelity for Circuit S increasing from 0.1724 to 0.5916 (a 243.2\% gain using a consistent $\epsilon_{\text{CNOT}}=0.05$ for evaluation). Similar trends of significant SWAP reduction are observed for larger circuits, strongly suggesting marked fidelity improvements for those as well, while incurring a modest increase in partitioning time of approximately 8-13\% compared to the \texttt{QuickPartitioner} workflow.

This paper presents \texttt{Fidelipart}’s methodology, implementation, and experimental evaluation. Section~\ref{sec:related} reviews related work in quantum circuit partitioning. Section~\ref{sec:method} details the proposed method, including hypergraph modeling, partitioning, merging, and dependency analysis. Section~\ref{sec:results_main} evaluates \texttt{Fidelipart} against \texttt{QuickPartitioner} across diverse circuit scenarios, Section~\ref{sec:discussion_main} interprets these findings, and Section~\ref{sec:conclusion_main} summarizes contributions and future directions. \texttt{Fidelipart} advances efficient, high-fidelity quantum circuit partitioning, enabling scalable compilation for NISQ-era devices.

\section{Related work}
\label{sec:related}
Quantum circuit partitioning is a cornerstone of quantum compilation, enabling the decomposition of large circuits into subcircuits to manage resource constraints, mitigate errors, and facilitate parallel execution on Noisy Intermediate-Scale Quantum (NISQ) devices \cite{nielsen2010, preskill2018}. Partitioning methods must balance gate counts, minimize cross-partition dependencies (cut qubits), and reduce communication overhead (e.g., SWAP gates) to preserve fidelity \cite{murali2019, li2019}. Below, we review existing approaches, focusing on BQSKit’s QuickPartitioner, hypergraph-based methods, and other compilation frameworks, highlighting their limitations and contrasting them with \texttt{Fidelipart}’s fidelity-aware hypergraph partitioning approach.

\subsection{BQSKit QuickPartitioner}
The Berkeley Quantum Synthesis Toolkit (BQSKit) includes QuickPartitioner, a heuristic-based method that groups gates into subcircuits based on qubit activity, such as consecutive gates acting on the same qubit pair \cite{bqskit2021}. While computationally efficient, QuickPartitioner often produces unbalanced partitions, leading to high cross-partition dependencies. For example, in a 24-qubit, 88-gate circuit, it generates 14 partitions with 23 cut qubits and 44 SWAP gates, resulting in low fidelity (e.g., 0.124147 when evaluated with $\epsilon_{\text{CNOT}}=0.05$) due to error accumulation from communication overhead \cite{bqskit2021}. Additionally, its partitioning time (e.g., 4.159 seconds for large circuits) limits scalability for complex circuits. Unlike QuickPartitioner, \texttt{Fidelipart} employs fidelity-aware hypergraph partitioning\cite{schlag2021} with Mt-KaHyPar \cite{Heuer2021mtkahypar}, using error-based hyperedge weights to minimize cut qubits and SWAP gates, achieving balanced partitions but has slightly longer partitioning times.

\subsection{Hypergraph-Based Partitioning}
Hypergraph partitioning has emerged as a powerful approach for capturing the complex connectivity of quantum circuits. Andrés-Martínez and Heunen \cite{andres2019} proposed a hypergraph-based method for distributing quantum circuits across multiple devices, using partitioning to minimize communication overhead during execution. Their approach optimizes for runtime interactions but does not incorporate quantum-specific optimizations, such as weighting hyperedges by gate error rates, leading to suboptimal partitioning for NISQ devices where high-error gates (e.g., CNOTs, with $\epsilon_{\text{CNOT}} \approx 0.05$) significantly impact fidelity. Similarly, classical hypergraph partitioning tools like KaHyPar \cite{schlag2021} excel at minimizing cut hyperedges but do not account for quantum gate error rates or qubit lifetimes, limiting their effectiveness for fidelity-critical applications. In contrast, \texttt{Fidelipart} models circuits as hypergraphs with hyperedges weighted by gate error rates (e.g., a CNOT with $\epsilon_{\text{CNOT}} = 0.05$ contributes to significantly higher hyperedge weights than single-qubit gates with $\epsilon_{\text{H}} = 0.001$). This fidelity-aware weighting, combined with partition merging based on shared qubit thresholds and dependency graph analysis for execution scheduling, ensures partitions prioritize fidelity while maintaining balanced gate distributions and efficient compilation. The specific strategy behind the relative weighting of different hyperedge types is further discussed in Section~\ref{sec:discussion_main}.

\texttt{Fidelipart} builds on these prior works by combining state-of-the-art hypergraph partitioning with quantum-specific optimizations, including error-based weighting and partition merging based on shared qubit thresholds. Its integration with BQSKit\cite{bqskit2021} and focus on fidelity make it a versatile solution for efficient quantum circuit compilation.

\subsection{Other Compilation Frameworks}
Other quantum compilation frameworks, such as Qiskit \cite{qiskit2019} and ScaffCC \cite{javadi2014}, also incorporate partitioning. Qiskit’s transpiler uses heuristic-based qubit mapping and routing to adapt circuits to hardware topologies, but its local optimization strategies often introduce excessive SWAP operations, increasing circuit depth and error rates \cite{li2019}. ScaffCC focuses on scalable compilation for large circuits but struggles with cross-partition dependencies, requiring costly stitching operations (e.g., ancilla qubits or teleportation) \cite{javadi2014,murali2019}. In contrast, \texttt{Fidelipart} integrates fidelity-aware hypergraph partitioning into BQSKit, leveraging global optimization to reduce cut qubits and SWAP gates while supporting scalable compilation for NISQ devices.

\section{Proposed method}
\label{sec:method}
Quantum circuit partitioning is vital for compiling algorithms onto Noisy Intermediate-Scale Quantum (NISQ) devices, limited by high gate errors and sparse qubit connectivity \cite{preskill2018,murali2019}. \texttt{Fidelipart} transforms quantum circuits into \emph{fidelity-aware hypergraphs} for optimized partitioning, explicitly designed to mitigate hardware noise. Unlike standard graph models limited to pairwise connections, it employs a hypergraph where gates become nodes. Crucially, it defines two hyperedge types: one representing the direct involvement of qubits in a multi-qubit gate (\emph{spatial dependencies}) and another grouping gates along individual qubit operational sequences (\emph{temporal dependencies}) \cite{schlag2021}. This dual representation enables precise, fidelity-aware modeling. We assign weights to both nodes and hyperedges, derived inversely from gate error rates and structural properties. This strongly penalizes cuts through operations involving high-error gates (e.g., CNOT, with an assumed $\epsilon_{\text{CNOT}} = 0.05$ for weighting) over robust ones (e.g., Hadamard, $\epsilon_{\text{H}} = 0.001$), guiding the partitioner. The resulting weighted hypergraph is then processed by a state-of-the-art partitioner (like Mt-KaHyPar) using the $k-1$ metric, aiming to minimize the total weight of cut hyperedges while balancing partition sizes. This minimizes not only \emph{cut qubits} but specifically *high-error* cuts, leading to a significant reduction in SWAP gate insertions (estimated logical realignments) and enhancing overall circuit fidelity beyond methods like QuickPartitioner \cite{bqskit2021}.

Under the \emph{Linear Chain Assumption}, qubits form a one-dimensional topology, requiring SWAP gates to connect non-adjacent qubits, which increases depth and errors ($\epsilon_{\text{SWAP}} \approx 0.03$, where each SWAP is often decomposed into three CNOT-like operations) \cite{li2019}. For instance, connecting distant qubits may need multiple SWAPs, worsening fidelity. \texttt{Fidelipart} minimizes cut qubits to reduce SWAPs and prioritizes low-error cuts, unlike heuristic methods that cut high-error gates. This framework is extensible to generalized topologies (e.g., 2D grids) for future quantum hardware.

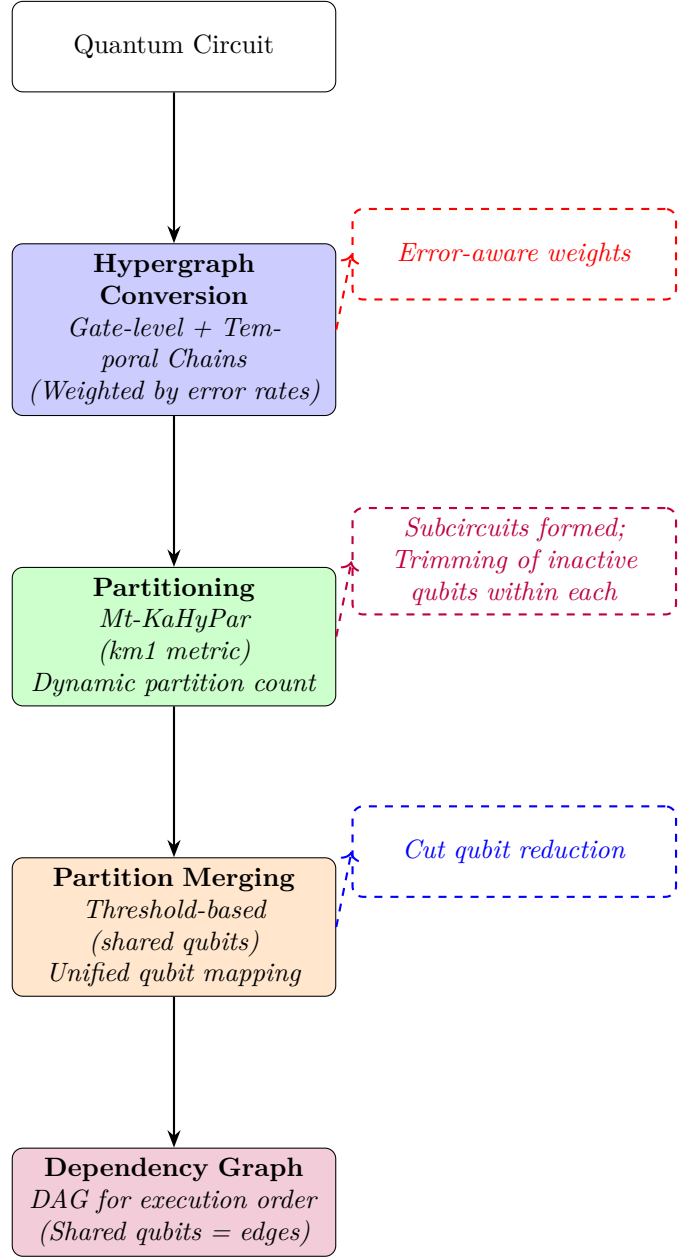
\begin{figure}[htbp!]
\centering
\begin{tikzpicture}[
    node distance=2cm,
    every node/.style={draw, rectangle, rounded corners, minimum height=1.2cm, text width=4cm, align=center, font=\small},
    arrow/.style={-Stealth, thick},
    hypergraph/.style={fill=blue!20},
    partitioning/.style={fill=green!20},
    merging/.style={fill=orange!20},
    dependency/.style={fill=purple!20}
]
\node (circuit) {Quantum Circuit};
\node (hypergraph) [below=of circuit, hypergraph] {
    \textbf{Hypergraph Conversion} \\
    \textit{Gate-level + Temporal Chains} \\
    \textit{(Weighted by error rates)}
};
\node (partitioning) [below=of hypergraph, partitioning] {
    \textbf{Partitioning} \\
    \textit{Mt-KaHyPar (km1 metric)} \\
    \textit{Dynamic partition count}
};
\node (merging) [below=of partitioning, merging] {
    \textbf{Partition Merging} \\
    \textit{Threshold-based (shared qubits)} \\
    \textit{Unified qubit mapping}
};
\node (dependency) [below=of merging, dependency] {
    \textbf{Dependency Graph} \\
    \textit{DAG for execution order} \\
    \textit{(Shared qubits = edges)}
};
\draw[arrow] (circuit) -- (hypergraph);
\draw[arrow] (hypergraph) -- (partitioning);
\draw[arrow] (partitioning) -- (merging);
\draw[arrow] (merging) -- (dependency);
\draw[->, thick, dashed, red] (hypergraph.east) -- ++(0.2cm,1cm)
    node[right, align=center] {\textcolor{red}{\textit{Error-aware weights}}};
\draw[->, thick, dashed, purple] (partitioning.east) -- ++(0.2cm,1cm) 
    node[right, align=center] {\textcolor{purple}{\textit{Subcircuits formed; Trimming of inactive qubits within each}}}; 
\draw[->, thick, dashed, blue] (merging.east) -- ++(0.2cm,1cm)
    node[right, align=center] {\textcolor{blue}{\textit{Cut qubit reduction}}};
\end{tikzpicture}
\caption{Workflow of the \texttt{Fidelipart} quantum circuit partitioning framework. 
The input quantum circuit is first transformed into a fidelity-aware hypergraph, where nodes represent gates and hyperedges (capturing multi-qubit interactions and temporal dependencies) are assigned \textit{error-aware weights} that penalize high-error operations. 
This weighted hypergraph is then partitioned using Mt-KaHyPar, optimizing the \textit{km1 metric} to minimize inter-partition connectivity. 
Following this, initial subcircuits are formed from the partitioned gates, and a crucial \textit{trimming} step identifies active global qubits within each partition, remapping them to local contiguous indices and removing unused qubits from each subcircuit's representation. 
Subsequently, an optional \textit{partition merging} stage can further reduce \textit{cut qubit} overhead by combining partitions that share a significant number of global qubits. 
Finally, a dependency graph (DAG) is constructed to define the valid execution order based on remaining shared qubits between the finalized partitions.}
\label{fig:workflow} 
\end{figure}

\texttt{Fidelipart}’s pipeline integrates four components: hypergraph conversion, partitioning, merging, and dependency graph construction, as shown in Figure~\ref{fig:workflow}. Like an optimized factory assembly line, it assigns gates (tasks) to partitions (workstations) to minimize worker movement (SWAPs) and ensure quality (fidelity). Poor partitioning, akin to inefficient task assignments, increases SWAPs; \texttt{Fidelipart}’s fidelity-aware approach yields robust subcircuits for NISQ devices.

\vspace{-0.7cm}

\subsection{Rationale and Objectives}
\label{subsec:rationale}
Quantum circuit partitioning must minimize high-error gate cuts while preserving gate execution order and qubit connectivity under the Linear Chain topology, where SWAP gates ($\epsilon_{\text{SWAP}} \approx 0.03$) enable non-adjacent qubit interactions \cite{li2019}. Existing methods like QuickPartitioner fail to model multi-gate dependencies or penalize high-error cuts \cite{bqskit2021}. Standard partitioning often ignores CNOT error rates (assumed $\epsilon_{\text{CNOT}} = 0.05$ in our model), leading to cuts that effectively increase error due to subsequent SWAP insertions needed to bridge these cuts, degrading fidelity. These gaps necessitate an algorithm tailored to fidelity-aware partitioning.

\texttt{Fidelipart} addresses this by converting circuits into hypergraphs with dual hyperedges—gate-level for multi-qubit gates (e.g., CNOTs, contributing to hyperedge weights like $100 \times 2 \times (1/0.05) = 4,000$) and temporal chains for qubit timelines—weighted by error rates (e.g., CNOTs are considered significantly more error-prone than Hadamards with $\epsilon_{\text{H}} = 0.001$) \cite{schlag2021}. Using Mt-KaHyPar’s km1 metric (5\% imbalance), it minimizes cut hyperedge weights, reducing high-error cuts and SWAPs \cite{schlag2021}. \texttt{Fidelipart}’s pipeline ensures robust NISQ execution. Its objectives are:
\begin{itemize}
  \item Minimize km1 metric cuts, reducing situations that necessitate cutting CNOT operations (contributing significantly to error, e.g., $\epsilon_{\text{CNOT}} = 0.05$) to enhance fidelity.
  \item Limit cut qubits to reduce SWAP gates (estimated logical realignments) in Linear Chain topology.
  \item Group gates in temporal chain hyperedges to preserve execution order and data locality.
  \item Model multi-qubit gates via specific hyperedge considerations to maintain connectivity.
  \item Balance partitions within 5\% for efficient NISQ execution.
\end{itemize}

\subsection{Core Concept: Fidelity-Aware Hypergraphs}
\label{subsec:hypergraphs}
\subsubsection{Why Hypergraphs?}
Traditional graph-based models for quantum circuit partitioning rely on pairwise edges to represent two-qubit interactions, limiting their ability to capture complex circuit dependencies \cite{andres2019}. For instance, a graph edge can model a CNOT gate between two qubits but cannot directly represent general multi-qubit gates or the temporal sequence of gates on a single qubit as a unified weighted entity for partitioning. Similarly, heuristic methods like QuickPartitioner fail to model these dependencies comprehensively or prioritize gate error rates, leading to suboptimal partitions with increased cut qubits and SWAP gates, degrading fidelity under the Linear Chain topology \cite{bqskit2021,li2019}. Hypergraphs, using hyperedges—sets of arbitrary nodes—offer a transformative advantage for Noisy Intermediate-Scale Quantum (NISQ) circuit partitioning \cite{schlag2021}.

In \texttt{Fidelipart}, hypergraphs enable precise modeling of \emph{multi-qubit interactions} and \emph{qubit timelines}. For multi-qubit gates, rather than a direct hyperedge connection between qubit nodes (which are not primary nodes in our gate-centric hypergraph), the gate node itself is part of a conceptual hyperedge whose weight reflects the cost of cutting this interaction. For qubit timelines, a single hyperedge directly groups all gate nodes acting on a specific qubit, preserving temporal dependencies. This flexibility aims to reduce the number of cut qubits—qubits split across partitions—thereby minimizing the need for SWAP gates, which incur significant errors ($\epsilon_{\text{SWAP}} \approx 0.03$). By capturing these dependencies with appropriate weighting, hypergraphs ensure partitions that respect circuit structure, enhancing fidelity over graph-based methods and enabling robust execution on NISQ devices.

\subsubsection{Representing Quantum Circuits}
\texttt{Fidelipart} represents a quantum circuit as a hypergraph where each node corresponds to a gate, such as a CNOT or Hadamard, directly reflecting the circuit’s operations. Hyperedges are defined in two distinct types to capture the circuit’s structure comprehensively:

\begin{itemize}
  \item \emph{Spatial Hyperedges (for Multi-Qubit Gates)}: These represent the structural integrity of multi-qubit gates, such as CNOTs or Toffoli gates, which enforce qubit connectivity. For a multi-qubit gate acting on $k$ qubits (represented by node $i$), \texttt{Fidelipart} creates a hyperedge containing only the gate’s node, i.e., $\{i\}$. The weight of this hyperedge (detailed in Sec 3.2.3) is high, penalizing any partition cut that would effectively "split" this gate's execution context across different partitions, thus indirectly protecting the multi-qubit interaction it embodies. For example, a CNOT gate (node $i$) forms such a hyperedge $\{i\}$; cutting this hyperedge in the partitioning process means gate $i$ is assigned to a boundary, likely increasing communication (SWAP) cost to implement its action across partitions.
  \item \emph{Temporal Hyperedges (for Qubit Timelines)}: These capture the execution order of gates along a qubit’s timeline. For each qubit $q$, \texttt{Fidelipart} constructs a hyperedge containing the indices of all gates operating on $q$, denoted $\{i_1, i_2, \ldots, i_m\}$, where $m$ is the number of gates. For instance, if a qubit undergoes a Hadamard, a CNOT, and a Pauli-X gate, the hyperedge includes all three gate nodes. This ensures that temporal dependencies—critical for correct circuit execution and maintaining coherent qubit evolution—are preserved during partitioning. The significant weight assigned to these hyperedges (see Sec 3.2.3) underscores the importance of temporal locality.
\end{itemize}

This dual hyperedge representation, rooted in the circuit’s operational and temporal structure, allows \texttt{Fidelipart} to model quantum circuits with unprecedented fidelity, addressing both spatial constraints (qubit connectivity, via penalties on splitting gate contexts) and temporal constraints (gate order) in a unified framework.

\subsubsection{Fidelity-Aware Weighting}
The cornerstone of \texttt{Fidelipart}’s approach is its fidelity-aware weighting scheme, which prioritizes cuts through low-error gates to mitigate the impact of NISQ hardware noise. Both nodes and hyperedges are assigned weights. Node weights are inversely proportional to their gate error rates. Hyperedge weights are designed to penalize partitions that split operations crucial for fidelity, such as those involving CNOT gates (with an assumed $\epsilon_{\text{CNOT}} = 0.05$ for weighting purposes) over robust ones, like Hadamards (with $\epsilon_{\text{H}} = 0.001$).

For nodes, each gate is weighted based on its error rate. A CNOT gate receives a weight of $10 \times (1/\epsilon_{\text{CNOT}}) = 10 \times (1/0.05) = 200$, reflecting a 10x multiplier to emphasize its criticality. A Hadamard gate is weighted as $1 \times (1/\epsilon_{\text{H}}) = 1 \times (1/0.001) = 1000$. Hyperedge weights amplify this fidelity focus:

\begin{itemize}
  \item \emph{Spatial Hyperedge Weights}: For a multi-qubit gate (node $i$) on $k$ qubits, its associated hyperedge $\{i\}$ receives a weight of $100 \times k \times (1/\text{gate\_error\_rate})$. For a CNOT gate ($k=2$, $\epsilon_{\text{CNOT}} = 0.05$), the weight is $100 \times 2 \times (1/0.05) = 4,000$. This high cost strongly discourages assigning the gate node $i$ to a partition boundary in a way that would necessitate splitting its operational context across partitions.
  \item \emph{Temporal Hyperedge Weights}: For a qubit with $m$ gates, the hyperedge (grouping these $m$ gate nodes) weight is $100 \times \lfloor m/2 \rfloor \times (1/\epsilon_{\text{H}})$. For example, a qubit with 4 gates yields a weight of $100 \times (4//2) \times (1/0.001) = 200,000$. The use of $\epsilon_H$ (a low error rate) as the divisor results in very high weights for temporal chains, strongly prioritizing keeping gates on the same qubit grouped to preserve execution order and data locality. The strategic decision behind the significant weight of temporal hyperedges is discussed further in Section~\ref{sec:discussion_main}.
\end{itemize}
To ensure numerical stability for Mt-KaHyPar\cite{Heuer2021mtkahypar}, hyperedge weights are normalized by scaling each weight as $w \times 10^6 / \max(w)$, maintaining relative penalties. The km1 metric, employed by Mt-KaHyPar with a 5\% imbalance allowance, minimizes the total weight of cut hyperedges, thus being guided by these error-informed weights \cite{schlag2021}. This weighting scheme ensures that \texttt{Fidelipart} produces partitions that aim to maximize circuit fidelity, reducing the impact of high-error operations and subsequent SWAP gates in NISQ execution.

\subsection{The Fidelipart Pipeline Architecture}
\label{subsec:pipeline}

\texttt{Fidelipart} employs a multi-stage pipeline to transform quantum circuits into fidelity-preserving partitions for Noisy Intermediate-Scale Quantum (NISQ) execution under the Linear Chain topology, as illustrated in Figure~\ref{fig:workflow}. The pipeline ensures circuits are divided into executable subcircuits that minimize high-error gate cuts and SWAP gate insertions, preserving circuit integrity \cite{li2019,schlag2021}. It comprises five stages: conversion, partitioning, reconstruction and trimming, merging, and dependency analysis.

The \emph{conversion} stage models the circuit as a weighted hypergraph, capturing gate dependencies and error rates (e.g., $\epsilon_{\text{CNOT}} = 0.05$, $\epsilon_{\text{H}} = 0.001$) to prioritize fidelity. \emph{Partitioning} divides this hypergraph into balanced subcircuits, minimizing cuts through error-prone gates to reduce error accumulation. The \emph{reconstruction and trimming} stage constructs initial subcircuits from the partitioned gates, optimizing qubit mappings for efficient execution by identifying active qubits and remapping them to local contiguous indices. \emph{Merging} refines these subcircuits by combining those sharing qubits, reducing communication overhead. Finally, \emph{dependency analysis} establishes the execution order across subcircuits, ensuring correctness. This pipeline produces partitions that enhance fidelity by limiting SWAP gates ($\epsilon_{\text{SWAP}} \approx 0.03$) and maintaining circuit structure, enabling robust NISQ performance.

\subsubsection{Conversion: circuit to hypergraph}
The conversion stage in \texttt{Fidelipart} transforms a quantum circuit into a weighted hypergraph, a pivotal step that enables fidelity-aware partitioning for Noisy Intermediate-Scale Quantum (NISQ) devices \cite{schlag2021}. Each gate, such as a Hadamard or CNOT, becomes a node in the hypergraph, capturing the circuit’s operational structure. Dependencies are modeled through two types of hyperedges: \emph{spatial hyperedges} (each containing a single multi-qubit gate node, heavily weighted to penalize splitting its context) encapsulate multi-qubit gate connectivity, while \emph{temporal hyperedges} group all gates acting on a single qubit, preserving execution order. For instance, in a 5-qubit circuit with 10 gates (5 Hadamards and 5 CNOTs), the hypergraph represents each CNOT’s connectivity (via its heavily weighted node-as-hyperedge) and each qubit’s gate sequence, ensuring comprehensive dependency modeling.

This hypergraph structure is critical because it encodes gate error rates, such as an assumed \(\epsilon_{\text{CNOT}} = 0.05\) versus \(\epsilon_{\text{H}} = 0.001\), into node and hyperedge weights, prioritizing low-error partitions \cite{andres2019}. High weights associated with error-prone gates, like CNOTs, discourage cuts that would introduce SWAP gates (\(\epsilon_{\text{SWAP}} = 0.03\)) in the Linear Chain topology, minimizing fidelity loss. By capturing both spatial and temporal relationships, the conversion stage ensures that subsequent partitioning respects circuit integrity, reducing error accumulation.

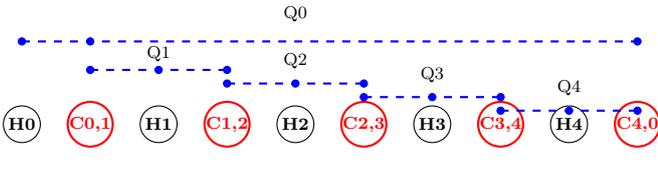
\begin{figure}[h]
\centering
\begin{tikzpicture}[
  node/.style={circle, draw, minimum size=0.6cm, inner sep=0pt, font=\scriptsize\bfseries},
  spatial_node/.style={red, thick, shape=circle, draw, minimum size=0.6cm, inner sep=0pt, font=\scriptsize\bfseries}, 
  temporal/.style={blue, dashed, thick},
  label/.style={font=\scriptsize},
  scale=0.9, transform shape,
  every node/.append style={scale=0.9}
]
\path[use as bounding box] (-4.25, -1.5) rectangle (4.25, 1.5);
\node[node] (H1) at (-4.5, 0) {H0};
\node[spatial_node] (C1) at (-3.5, 0) {C0,1}; 
\node[node] (H2) at (-2.5, 0) {H1};
\node[spatial_node] (C2) at (-1.5, 0) {C1,2};
\node[node] (H3) at (-0.5, 0) {H2};
\node[spatial_node] (C3) at (0.5, 0) {C2,3};
\node[node] (H4) at (1.5, 0) {H3};
\node[spatial_node] (C4) at (2.5, 0) {C3,4};
\node[node] (H5) at (3.5, 0) {H4};
\node[spatial_node] (C5) at (4.5, 0) {C4,0};
\draw[temporal] (-4.5, 1.22) -- (4.5, 1.22);
\filldraw[blue] (-4.5, 1.22) circle (1.5pt) (-3.5, 1.22) circle (1.5pt) (4.5, 1.22) circle (1.5pt);
\node[label, above=0.1cm] at (-0.5, 1.3) {Q0};
\draw[temporal] (-3.5, 0.8) -- (-1.5, 0.8);
\filldraw[blue] (-3.5, 0.8) circle (1.5pt) (-2.5, 0.8) circle (1.5pt) (-1.5, 0.8) circle (1.5pt);
\node[label, above=0.1cm] at (-2.5, 0.7) {Q1};
\draw[temporal] (-1.5, 0.6) -- (0.5, 0.6);
\filldraw[blue] (-1.5, 0.6) circle (1.5pt) (-0.5, 0.6) circle (1.5pt) (0.5, 0.6) circle (1.5pt);
\node[label, above=0.1cm] at (-0.5, 0.6) {Q2};
\draw[temporal] (0.5, 0.4) -- (2.5, 0.4);
\filldraw[blue] (0.5, 0.4) circle (1.5pt) (1.5, 0.4) circle (1.5pt) (2.5, 0.4) circle (1.5pt);
\node[label, above=0.1cm] at (1.5, 0.4) {Q3};
\draw[temporal] (2.5, 0.2) -- (4.5, 0.2);
\filldraw[blue] (2.5, 0.2) circle (1.5pt) (3.5, 0.2) circle (1.5pt) (4.5, 0.2) circle (1.5pt);
\node[label, above=0.1cm] at (3.5, 0.2) {Q4};
\end{tikzpicture}
\caption{Hypergraph representation of a 5-qubit circuit with 10 gates (Hadamards H0--H4, CNOTs C0,1--C4,0). CNOT gate nodes (red circles, associated with spatial hyperedge weight 4,000 each) represent points of high penalty if their operational context is split. Temporal hyperedges (blue dashed lines, e.g., for Q0 with weight $\approx 100,000 \times \lfloor 3/2 \rfloor = 100,000$) connect gates per qubit timeline, with dots marking gate positions (e.g., Q0: H0, C0,1, C4,0). Node weights (1000 for Hadamards, 200 for CNOTs) reflect error rates (\(\epsilon_{\text{CNOT}} = 0.05\), \(\epsilon_{\text{H}} = 0.001\)), guiding fidelity-aware partitioning.}
\label{fig:hypergraph}
\end{figure}

Figure~\ref{fig:hypergraph} illustrates this process for the example circuit, depicting nodes for each gate, conceptual spatial hyperedges (represented by the CNOT gate nodes themselves having properties that lead to high cut costs), and explicit temporal hyperedges for qubit timelines. This robust representation underpins \texttt{Fidelipart}’s ability to produce partitions that enhance circuit fidelity and enable reliable NISQ execution, distinguishing it from traditional graph-based approaches \cite{andres2019}.

\subsubsection{Partitioning: kahypar partitioning}
The partitioning stage in \texttt{Fidelipart} divides the weighted hypergraph into subcircuits using an advanced external solver, Mt-KaHyPar\cite{Heuer2021mtkahypar}, to create executable partitions for Noisy Intermediate-Scale Quantum (NISQ) devices \cite{schlag2021}. It assigns each gate (node) to a partition, guided by hyperedge weights that reflect gate error rates and connectivity. For a 5-qubit circuit with 10 gates, the stage splits nodes into two balanced subcircuits, ensuring minimal cuts through high-weight hyperedges, such as those representing CNOT gate contexts or qubit timelines. This process leverages a sophisticated metric to prioritize low-error gate assignments while maintaining roughly equal partition sizes, within a 5\% imbalance tolerance.

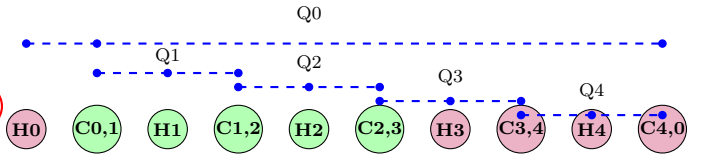
\begin{figure}[h]
\centering
\begin{tikzpicture}[
  node/.style={circle, draw, minimum size=0.6cm, inner sep=0pt, font=\scriptsize\bfseries},
  spatial_node/.style={shape=circle, draw, minimum size=0.6cm, inner sep=0pt, font=\scriptsize\bfseries}, 
  temporal/.style={blue, dashed, thick},
  label/.style={font=\scriptsize},
  scale=0.93, transform shape,
  every node/.append style={scale=0.93}
]

\path[use as bounding box] (-4.25, -1.5) rectangle (4.25, 1.5);

\node[node, fill=purple!30] (H1) at (-4.5, 0) {H0};
\node[spatial_node, fill=green!30] (C1) at (-3.5, 0) {C0,1}; 
\node[node, fill=green!30] (H2) at (-2.5, 0) {H1};
\node[spatial_node, fill=green!30] (C2) at (-1.5, 0) {C1,2};
\node[node, fill=green!30] (H3) at (-0.5, 0) {H2};
\node[spatial_node, fill=green!30] (C3) at (0.5, 0) {C2,3};
\node[node, fill=purple!30] (H4) at (1.5, 0) {H3};
\node[spatial_node, fill=purple!30] (C4) at (2.5, 0) {C3,4};
\node[node, fill=purple!30] (H5) at (3.5, 0) {H4};
\node[spatial_node, fill=purple!30] (C5) at (4.5, 0) {C4,0};

\draw[temporal] (-4.5, 1.22) -- (4.5, 1.22);
\filldraw[blue] (-4.5, 1.22) circle (1.5pt) (-3.5, 1.22) circle (1.5pt) (4.5, 1.22) circle (1.5pt);
\node[label, above=0.1cm] at (-0.5, 1.3) {Q0};
% Q1: C1, H2, C2
\draw[temporal] (-3.5, 0.8) -- (-1.5, 0.8);
\filldraw[blue] (-3.5, 0.8) circle (1.5pt) (-2.5, 0.8) circle (1.5pt) (-1.5, 0.8) circle (1.5pt);
\node[label, above=0.1cm] at (-2.5, 0.7) {Q1};
% Q2: C2, H3, C3
\draw[temporal] (-1.5, 0.6) -- (0.5, 0.6);
\filldraw[blue] (-1.5, 0.6) circle (1.5pt) (-0.5, 0.6) circle (1.5pt) (0.5, 0.6) circle (1.5pt);
\node[label, above=0.1cm] at (-0.5, 0.6) {Q2};
% Q3: C3, H4, C4
\draw[temporal] (0.5, 0.4) -- (2.5, 0.4);
\filldraw[blue] (0.5, 0.4) circle (1.5pt) (1.5, 0.4) circle (1.5pt) (2.5, 0.4) circle (1.5pt);
\node[label, above=0.1cm] at (1.5, 0.4) {Q3};
% Q4: C4, H5, C5
\draw[temporal] (2.5, 0.2) -- (4.5, 0.2);
\filldraw[blue] (2.5, 0.2) circle (1.5pt) (3.5, 0.2) circle (1.5pt) (4.5, 0.2) circle (1.5pt);
\node[label, above=0.1cm] at (3.5, 0.2) {Q4};

\end{tikzpicture}
\caption{Partitioned hypergraph for a 5-qubit circuit with 10 gates (Hadamards H0--H4, CNOTs C0,1--C4,0), divided into two subcircuits. Nodes in partition 1 (green: C0,1, H1, C1,2, H2, C2,3) and partition 0 (purple: H0, H3, C3,4, H4, C4,0) reflect the assignment by Mt-KaHyPar. Spatial hyperedges (associated with CNOT gate nodes, carrying significant weight, e.g., 4,000 before normalization) and temporal hyperedges (blue dashed lines, carrying very high weight, e.g., $\sim$200,000 for a 4-gate chain before normalization) connect qubit timelines with dots at gate positions (e.g., Q0: H0, C0,1, C4,0). Node weights (1000 for Hadamards, 200 for CNOTs) and error rates (\(\epsilon_{\text{CNOT}} = 0.05\), \(\epsilon_{\text{H}} = 0.001\)) guide partitioning to minimize SWAP gates (\(\epsilon_{\text{SWAP}} = 0.03\)), preserving fidelity.}
\label{fig:partitioned_hypergraph}
\end{figure}

Partitioning is crucial because it directly impacts circuit fidelity by minimizing the need for SWAP gates (\(\epsilon_{\text{SWAP}} = 0.03\)) in the Linear Chain topology, where cuts through multi-qubit gates introduce costly communication overhead \cite{andres2019}. By preserving spatial and temporal dependencies encoded in hyperedges, the stage ensures subcircuits respect the circuit’s structure, reducing error accumulation from high-error gates like CNOTs (\(\epsilon_{\text{CNOT}} = 0.05\)). Figure~\ref{fig:partitioned_hypergraph} illustrates this for the example circuit, showing nodes assigned to two partitions, with one group containing early gates (e.g., Hadamards and initial CNOTs) and another containing later gates. This strategic division enhances \texttt{Fidelipart}’s ability to deliver robust, high-fidelity partitions for NISQ execution.

\subsubsection{Reconstruction \& Trimming}
\label{subsubsec:reconstruction_trimming} 

The reconstruction and trimming stage in \texttt{Fidelipart} transforms partitioned gate assignments into executable subcircuits, a crucial step for efficient execution on Noisy Intermediate-Scale Quantum (NISQ) devices. This stage begins by grouping gates according to their assigned partition labels, thereby constructing an initial subcircuit for each partition (e.g., if the original circuit has 10 gates and is divided into two partitions, this step will initially delineate two groups of gates).

For each of these initial subcircuits, the stage then performs two key operations:
\begin{enumerate}
    \item \textbf{Active Qubit Identification:} It identifies the set of unique \textit{active global qubits}—those original circuit qubit indices that are actually involved in the gates assigned to that specific partition. Inactive qubits (those not acted upon by any gate in this particular collection) are effectively ignored for this subcircuit.
    \item \textbf{Local Contiguous Re-mapping:} These identified active global qubits are then \textbf{sorted} (based on their original numerical index) and subsequently mapped to a new, compact set of \textit{local physical indices} starting from 0. For example, if a partition's sorted active global qubits are \{0, 1, 3, 4\}, they would be remapped such that global qubit 0 maps to local physical qubit 0, global qubit 1 to local 1, global qubit 3 to local 2, and global qubit 4 to local 3 within that partition's context.
\end{enumerate}
Finally, a new, trimmed subcircuit is constructed where all gate operations use these new local physical indices. This entire process creates smaller circuits precisely tailored to the active qubits, significantly reducing resource demands while preserving the original gate operations as intended by the partitioning.

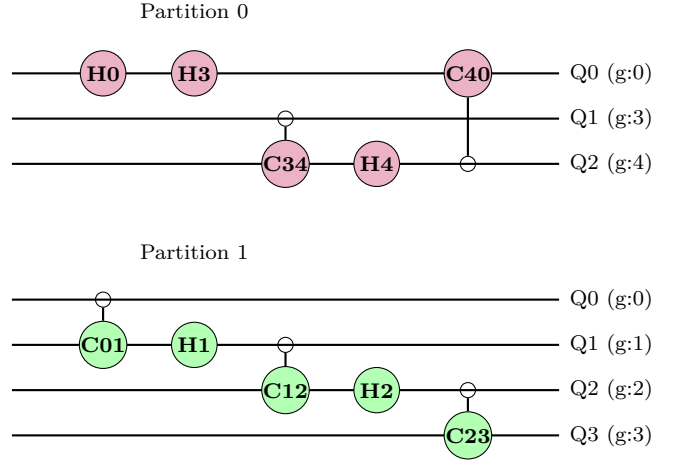
\begin{figure}[h]
\centering
\begin{tikzpicture}[
  gate/.style={circle, draw, minimum size=0.6cm, inner sep=0pt, font=\scriptsize\bfseries},
  gate0/.style={gate, fill=purple!30},
  gate1/.style={gate, fill=green!30},
  target/.style={circle, draw, minimum size=0.2cm, inner sep=0pt},
  line/.style={thick},
  label/.style={font=\scriptsize},
  scale=1.0, transform shape,
  every node/.append style={scale=1.0}
]

\path[use as bounding box] (-5, -4.5) rectangle (5, 2.5);

\node[label, above=0.1cm] at (-2.4, 2.3) {Partition 0};
% Qubit lines
\draw[line] (-4.8, 1.8) -- (2.4, 1.8) node[right, label] {Q0 (g:0)};
\draw[line] (-4.8, 1.2) -- (2.4, 1.2) node[right, label] {Q1 (g:3)};
\draw[line] (-4.8, 0.6) -- (2.4, 0.6) node[right, label] {Q2 (g:4)};
% Gates
\node[gate0] (H0) at (-3.6, 1.8) {H0};
\node[gate0] (H3_fig) at (-2.4, 1.8) {H3}; % Renamed node for clarity
\node[target] (C34_target) at (-1.2, 1.2) {}; 
\node[gate0] (C34) at (-1.2, 0.6) {C34}; 
\draw[line] (C34_target) -- (C34); 

\node[gate0] (H4_fig) at (0, 0.6) {H4}; % Renamed node for clarity
\node[target] (C40_target) at (1.2, 0.6) {}; 
\node[gate0] (C40) at (1.2, 1.8) {C40}; 
\draw[line] (C40_target) -- (C40);

% Partition 1 Subcircuit (4 qubits, 5 gates)
\node[label, above=0.1cm] at (-2.4, -0.9) {Partition 1};
% Qubit lines
\draw[line] (-4.8, -1.2) -- (2.4, -1.2) node[right, label] {Q0 (g:0)};
\draw[line] (-4.8, -1.8) -- (2.4, -1.8) node[right, label] {Q1 (g:1)};
\draw[line] (-4.8, -2.4) -- (2.4, -2.4) node[right, label] {Q2 (g:2)};
\draw[line] (-4.8, -3.0) -- (2.4, -3.0) node[right, label] {Q3 (g:3)};
% Gates
\node[target] (C01_target) at (-3.6, -1.2) {}; 
\node[gate1] (C01) at (-3.6, -1.8) {C01}; 
\draw[line] (C01_target) -- (C01); 
\node[gate1] (H1_fig) at (-2.4, -1.8) {H1}; % Renamed node for clarity

\node[target] (C12_target) at (-1.2, -1.8) {}; 
\node[gate1] (C12) at (-1.2, -2.4) {C12}; 
\draw[line] (C12_target) -- (C12); 

\node[gate1] (H2_fig) at (0, -2.4) {H2}; % Renamed node for clarity

\node[target] (C23_target) at (1.2, -2.4) {}; 
\node[gate1] (C23) at (1.2, -3.0) {C23}; 
\draw[line] (C23_target) -- (C23); 

\end{tikzpicture}
\caption{Subcircuits for a 5-qubit circuit with 10 gates, constructed from partition labels. Partition 0 (purple, top) includes 5 gates (H(0), H(3), CNOT(3,4), H(4), CNOT(4,0)) on 3 physical qubits (mapped from global qubits 0, 3, 4 to local Q0, Q1, Q2; map: $\{0: 0, 3: 1, 4: 2\}$). Partition 1 (green, bottom) includes 5 gates (CNOT(0,1), H(1), CNOT(1,2), H(2), CNOT(2,3)) on 4 physical qubits (mapped from global qubits 0, 1, 2, 3 to local Q0, Q1, Q2, Q3; map: $\{0: 0, 1: 1, 2: 2, 3: 3\}$). Active qubits are identified using global indices, then remapped to contiguous local physical indices, minimizing resources while preserving fidelity (assuming $\epsilon_{\text{CNOT}} = 0.05$, $\epsilon_{\text{H}} = 0.001$) and reducing SWAP gate overhead ($\epsilon_{\text{SWAP}} = 0.03$) in the Linear Chain topology.}
\label{fig:subcircuits}
\end{figure}

This reconstruction and trimming stage is vital because it produces subcircuits that are both directly executable (due to localized qubit indexing) and optimized for the NISQ environment. It inherently maintains fidelity by adhering to the fidelity-aware partitions generated earlier by the hypergraph-based approach \cite{schlag2021}. By remapping qubits as described, it minimizes qubit overhead. This is particularly critical in constrained topologies such as a Linear Chain, where interactions between non-adjacent logical qubits (that might now be further apart due to partitioning) would necessitate SWAP gates (with an assumed error rate, e.g., $\epsilon_{\text{SWAP}} \approx 0.03$). The process ensures that high-error gates, like CNOTs (e.g., $\epsilon_{\text{CNOT}} \approx 0.05$), remain intact within the newly defined subcircuits, thus upholding the error-minimizing intent of the partitioning. Figure~\ref{fig:subcircuits} illustrates this for an example circuit, showing two subcircuits: one with five gates on four (local) qubits, and another with five gates on three (local) qubits, each with their qubits remapped to support robust NISQ execution.

\subsubsection{Merging}
The merging stage in \texttt{Fidelipart} refines the initial subcircuits by combining those with shared qubits, creating a unified quantum circuit with an optimized qubit mapping. It identifies overlapping qubits between partitions—such as those connecting operations across subcircuits—and merges them when the number of shared qubits exceeds a threshold, producing a single circuit that integrates all relevant gates. For instance, two subcircuits with overlapping qubit usage are fused, and their mappings are updated to a continuous physical index space, reducing fragmentation.

This stage is crucial for large circuits because it minimizes the number of cut qubits—qubits that span multiple partitions—thereby decreasing the need for costly SWAP operations in the Linear Chain topology (assuming \(\epsilon_{\text{SWAP}} = 0.03\)). By consolidating related operations, it enhances execution efficiency on Noisy Intermediate-Scale Quantum (NISQ) devices while preserving fidelity, as high-error gates like CNOTs (with assumed \(\epsilon_{\text{CNOT}} = 0.05\)) remain intact within the merged structure. 

\subsubsection{Dependency Analysis}
The dependency analysis stage in \texttt{Fidelipart} constructs a directed acyclic graph (DAG) to establish the execution order of merged subcircuits, ensuring quantum state consistency across partitions. It examines the physical qubits shared between partitions—such as those linking operations—and creates directed edges to dictate the sequence, with lower-indexed partitions executing before higher-indexed ones when dependencies exist. For example, two subcircuits with overlapping qubits are ordered to prevent state interference, forming a structured execution plan.

This stage is vital because it preserves the integrity of quantum states on Noisy Intermediate-Scale Quantum (NISQ) devices, where improper sequencing can introduce errors in high-fidelity gates like CNOTs (with $\epsilon_{\text{CNOT}} = 0.05$). By enforcing a clear order—such as executing a partition operating on qubit 0 before one sharing it—it enables parallel execution of independent subcircuits, optimizing resource use and reducing latency. Additionally, the DAG guides further optimization passes, minimizing SWAP overhead in the Linear Chain topology, enhancing overall circuit efficiency. This structured approach transforms the merged partitions into a coherent execution framework, critical for robust NISQ performance.

Consider the dependency between two subcircuits: one with gates on qubits 0, 3, and 4, and another on 0, 1, 2, and 3. The shared qubits {0, 3} necessitate that the first executes before the second, as reflected in the DAG edge from Partition 0 to Partition 1. This ensures that quantum states are preserved across operations, aligning with \texttt{Fidelipart}’s goal of fidelity-aware optimization.

\subsection{Assumptions and Scope}
\label{subsec:assumptions}
\subsubsection{Hardware Model}
\texttt{Fidelipart} initially targets the Linear Chain topology as its hardware model, a structure where qubits are arranged sequentially, and only adjacent qubits can interact directly via two-qubit gates like CNOTs. This model underpins the pipeline’s design, from hypergraph construction to dependency analysis, ensuring that gate operations align with the physical constraints of the hardware. The Linear Chain assumption shapes how partitions are formed and merged, focusing on minimizing the need for SWAP gates to enable non-adjacent qubit interactions. The current SWAP cost estimation model (detailed in Section~\ref{subsubsec:performance_metrics_algo_main}) quantifies logical realignments and simplifies the actual SWAP chain length required for non-adjacent qubits on the linear chain; a more detailed routing analysis incorporating physical distances is a potential area for future refinement.

The choice of the Linear Chain is strategic for Noisy Intermediate-Scale Quantum (NISQ) devices, as it mirrors the connectivity of some early and contemporary quantum hardware platforms \cite{ibm2021, rigetti2021}. This choice focuses the optimization challenge on minimizing SWAP overhead, as non-adjacent interactions must be enabled by these error-prone operations (e.g., assuming $\epsilon_{\text{SWAP}} \approx 0.03$). \texttt{Fidelipart}'s design aims to better preserve the effective fidelity of crucial native gates, such as \texttt{CNOTGate}s (with an assumed error $\epsilon_{\text{CNOT}} = 0.05$ for weighting and evaluation) and \texttt{HGate}s (with an assumed error $\epsilon_{\text{H}} = 0.001$ ), by reducing their exposure to additional errors from excessive SWAPing. This focus enables \texttt{Fidelipart} to effectively manage qubit mappings and inform execution order considerations, contributing to higher-fidelity circuit execution in such constrained environments. By prioritizing a common NISQ-relevant topology, the pipeline addresses tangible hardware limitations while building a robust partitioning foundation.

However, \texttt{Fidelipart} is designed with extensibility in mind. The pipeline’s modular structure—particularly its hypergraph-based partitioning and dependency analysis—can adapt to other topologies, such as 2D grids or all-to-all connectivity, with adjustments to the partitioning strategy and SWAP insertion logic. This flexibility ensures that \texttt{Fidelipart} can evolve alongside advancements in quantum hardware, broadening its applicability for future NISQ systems.

\subsubsection{Solver Dependence: Mt-KaHyPar}
\label{subsubsec:solver_dependence} 

\texttt{Fidelipart} relies on Mt-KaHyPar\cite{Heuer2021mtkahypar}, a high-quality shared-memory hypergraph partitioning solver, to execute its core partitioning stage. This stage transforms the fidelity-aware hypergraph representation of a quantum circuit into a set of balanced subcircuits. Mt-KaHyPar\cite{Heuer2021mtkahypar} is configured to optimize the $k-1$ (\texttt{km1}) metric, which aims to minimize the total weight of cut hyperedges---effectively reducing the connectivity across partitions by minimizing the number of distinct partitions each hyperedge spans. Additionally, an imbalance constraint $\epsilon$ is enforced, which was set to 0.05 (5\%) in our configuration, to ensure that partitions remain relatively balanced in terms of their cumulative node weights, preventing any single subcircuit from becoming disproportionately large. The choice and configuration of Mt-KaHyPar directly influence \texttt{Fidelipart}'s ability to generate high-quality initial partitions suitable for subsequent merging and dependency analysis.

The use of Mt-KaHyPar\cite{Heuer2021mtkahypar} with the \texttt{km1} objective on our fidelity-weighted hypergraph aligns with \texttt{Fidelipart}’s goal of minimizing effective cut qubits. By penalizing cuts through hyperedges representing error-prone operations (e.g., \texttt{CNOTGate}s with an assumed $\epsilon_{\text{CNOT}} = 0.05$ ) or those crucial for temporal locality, the partitioning is guided towards solutions that inherently reduce the circumstances requiring SWAP gates (with an assumed $\epsilon_{\text{SWAP}} \approx 0.03$ ) in a constrained Linear Chain topology. Balanced partitions, as enforced by the imbalance parameter, further aim for equitable workload distribution, which can be beneficial for execution efficiency on NISQ devices. By leveraging Mt-KaHyPar’s\cite{Heuer2021mtkahypar} advanced multilevel partitioning capabilities, \texttt{Fidelipart} produces subcircuits that better maintain quantum state integrity while optimizing resource utilization.

However, this reliance on an external solver introduces a dependency that may affect portability or flexibility in environments where Mt-KaHyPar\cite{Heuer2021mtkahypar} is not readily available or compatible. Despite this consideration, Mt-KaHyPar’s established effectiveness and performance in hypergraph partitioning justify its integration, providing a robust foundation for \texttt{Fidelipart}’s fidelity-aware optimization strategy. Future work could explore alternative hypergraph partitioning solvers or investigate the development of custom, integrated partitioning heuristics tailored specifically for this quantum context to potentially reduce external dependencies while maintaining or enhancing performance.

\subsubsection{Gate Set and Error Assumptions}
\label{subsubsec:gate_set} 

\texttt{Fidelipart}'s fidelity-aware design primarily focuses on a core set of gates prevalent in Noisy Intermediate-Scale Quantum (NISQ) circuits: the Hadamard (\texttt{HGate}), \texttt{CNOTGate}, and \texttt{SwapGate}. The assumed error rates used to inform \texttt{Fidelipart}'s internal weighting schemes and for final fidelity evaluation are $\epsilon_{\text{H}} = 0.001$ for Hadamard, $\epsilon_{\text{CNOT}} = 0.05$ for CNOT, and $\epsilon_{\text{SWAP}} = 0.03$ for SWAP (where a SWAP is modeled as three CNOTs for error contribution purposes, making its effective single-operation error based on CNOTs even higher). These values, reflecting typical characterizations of some current NISQ hardware or chosen as representative high-error scenarios, provide a consistent foundation for the pipeline’s optimization strategy by enabling a quantitative assessment of the potential fidelity impact of different partitioning choices.

This assumed gate error landscape directly shapes \texttt{Fidelipart}’s optimization approach. The hypergraph construction and weighting strategy are designed to create partitions that inherently minimize the impact of higher-error operations. Specifically, it penalizes cuts through \texttt{CNOTGate}s and seeks to arrange partitions such that the subsequent need for error-prone \texttt{SwapGate}s (critical for enabling non-adjacent qubit interactions on a Linear Chain topology) is reduced. While these gates form the core focus for explicit fidelity-aware weighting, \texttt{Fidelipart}, by virtue of its integration within the BQSKIT framework\cite{bqskit2021}, can process circuits containing any gate type supported by BQSKIT\cite{bqskit2021}. This includes other single-qubit gates, multi-qubit gates, and three-qubit gates like the \texttt{CCXGate} (Toffoli gate), as demonstrated in our benchmark circuits (e.g., ``Circuit L''). These other gates are assigned weights based on available error rate information or default assumptions within the \texttt{circuit\_to\_hypergraph} process (Algorithm~\ref{algo:circuit_to_hypergrap_algo}). This flexibility ensures the pipeline’s broad applicability to diverse quantum circuits while maintaining its optimization focus on the dominant error sources.

The choice of H, CNOT, and SWAP as the primary gates for fidelity-aware optimization targets circuits commonly found in quantum algorithms, such as variational circuits, which are often dominated by single-qubit rotations and two-qubit entangling gates. By leveraging BQSKIT’s extensible gate model, \texttt{Fidelipart} is well-positioned to adapt to more complex gate sets and refined error models as quantum hardware and characterization techniques evolve, broadening its utility for future NISQ applications.

\subsubsection{Merge Heuristics}
\label{subsubsec:merge_heuristics} 

\texttt{Fidelipart} employs a greedy, heuristic approach in its optional Merging stage, designed to further refine partitions by combining subcircuits based on the number of shared global qubits between them. This strategy iteratively prioritizes merging pairs of partitions that exhibit the most significant qubit overlap—specifically, those sharing a number of global qubits at or above a specified \texttt{threshold}. For example, if two partitions were found to share global qubits $\{q_x, q_y\}$ and this met the threshold, they would be combined into a single, larger subcircuit. This process involves unifying their respective \texttt{\{global:local\}} qubit mappings into a new, consolidated map for the merged entity, and reconstructing the circuit with operations from both original partitions, now using the new unified local qubit indices. An example of this merge operation was demonstrated in the illustrative walkthrough (Section~\ref{para:walkthrough_merging_results}).

The choice of a greedy heuristic for merging is particularly well-suited for NISQ-era optimization, where computational efficiency in the compilation pipeline is critical. By focusing on the most beneficial merges first (i.e., those with the largest number of shared qubits), \texttt{Fidelipart} aims to quickly consolidate strongly related clusters of operations. This, in turn, is intended to reduce the number of cut qubits, thereby minimizing the subsequent need for SWAP operations (with an assumed $\epsilon_{\text{SWAP}} \approx 0.03$ ) when mapped to a Linear Chain topology. Such reductions contribute to preserving the fidelity of high-error gates like \texttt{CNOTGate}s (with an assumed $\epsilon_{\text{CNOT}} = 0.05$) by minimizing their exposure to additional error sources. This merging approach leverages the high-quality initial partitions produced by Mt-KaHyPar\cite{Heuer2021mtkahypar}, ensuring that the refinement process builds upon a solid foundation while further optimizing for practical execution constraints.

However, the greedy nature of this heuristic means that it may converge to locally optimal solutions, potentially missing a globally optimal partition configuration that could offer even fewer cuts. While effective for the targeted NISQ circuits and providing a balance between optimization quality and runtime, this inherent trade-off suggests opportunities for future refinement. Alternative, more computationally intensive strategies, such as those based on simulated annealing or reinforcement learning, could be explored to potentially achieve improved cut minimization, particularly for larger or more complex quantum circuits, thereby further enhancing \texttt{Fidelipart}’s performance.

\section{Algorithm Design and Details}
\label{sec:algorithm}
\subsection{Hypergraph Conversion (circuit\_to\_hypergraph)}
\label{subsec:hypergraph_conversion_algo} 

\subsubsection{Input/Output}
The \texttt{circuit\_to\_hypergraph} function takes a quantum circuit object (\texttt{circuit}) containing a list of gates and their qubit targets, an output file path (\texttt{output\_file}, default "circuit.hgr") for the hypergraph representation, and an optional dictionary \texttt{gate\_error\_rates} mapping gate names to their error rates (e.g., \{"CNOTGate": 0.05, "HGate": 0.001, "SWAP": 0.03\}). It produces no direct return value, instead writing the hypergraph in hMETIS format to the specified file, enabling downstream partitioning.

\subsubsection{Node Weighting}
Node weights are assigned based on gate criticality, using inverse error rates to prioritize low-error gates. For CNOT gates, a weight of \(10 \times (1 / \epsilon_{\text{CNOT}})\) (where $\epsilon_{\text{CNOT}} = 0.05$) is applied, reflecting their 10x importance due to higher error rates. Other gates, such as Hadamard, receive \(1 \times (1 / \epsilon_{\text{gate}})\), with defaults of \(1 \times (1 / 0.001)\) for HGate and \(1 \times (1 / 0.01)\) for others if unspecified. This 10x factor for CNOT gates emphasizes preserving multi-qubit interactions, critical in the Linear Chain topology.

\subsubsection{Hyperedge Construction}
Hyperedges are constructed in two categories. \textbf{Gate-level hyperedges} capture multi-qubit interactions: for each gate acting on more than one qubit (e.g., CNOT), a hyperedge includes only that gate’s index, with a weight of \(100 \times \text{num\_qubits} \times (1 / \epsilon_{\text{gate}})\) (using $\epsilon_{\text{CNOT}} = 0.05$ for CNOTs). This penalizes cutting entangling gates, encouraging their inclusion in the same partition. \textbf{Temporal hyperedges} model qubit timelines: for each qubit, all gate indices acting on it are grouped into a hyperedge if more than one gate exists. The weight is \(\max(1, 100 \times (\text{num\_gates\_on\_qubit} // 2) \times (1 / \epsilon_{\text{H}}))\), promoting consecutive operations on the same qubit to stay together, using Hadamard’s low error rate (\(\epsilon_{\text{H}} = 0.001\)) as a baseline. The significantly larger weights for temporal hyperedges reflect a strong prioritization of maintaining sequential operational integrity on individual qubits.

\subsubsection{Normalization}
Weights are normalized using a scaling factor of \(10^6\) to ensure compatibility with the hMETIS format, which requires integer weights. The process involves finding the maximum weight among valid edge weights, then computing \(\text{normalized\_weight} = (\text{weight} \times 10^6) / \max_{\text{weight}}\), rounded to the nearest integer. This scaling amplifies weight differences, enhancing partitioner sensitivity to error rates and temporal locality, while adhering to hMETIS constraints.

\subsubsection{Pseudocode}
\label{algo:circuit_to_hypergrap_algo} 
\paragraph{Input}
\begin{itemize}
    \item \texttt{circuit}: Quantum circuit object
    \item \texttt{output\_file}: String path for output file (default: "circuit.hgr")
    \item \texttt{gate\_error\_rates}: Dictionary mapping gate names to error rates (optional, e.g., \{"CNOTGate": 0.05, "HGate": 0.001\})
\end{itemize}
\paragraph{Output} \begin{itemize} \item None (writes to file) \end{itemize}
\paragraph{Algorithm Steps}
\textbf{1. INITIALIZATION}
\begin{lstlisting}[basicstyle=\ttfamily\scriptsize]
1.1 Import CNOTGate class for type checking
1.2 Extract list of gates/operations from circuit
1.3 Calculate:
    - num_gates = total number of gates
    - num_qubits = number of qubits in circuit
\end{lstlisting}
\textbf{2. NODE WEIGHT CALCULATION}
\begin{lstlisting}[numbers=none, basicstyle=\ttfamily\scriptsize] 
Initialize node_weights as an empty list
For each gate in circuit's operations:
    SET gate_name TO the name of the gate (e.g., "HGate", "CNOTGate")
    IF gate_name is "CNOTGate":
        error_rate_for_op = gate_error_rates.get("CNOTGate", 0.05) 
        node_weight = 10 * (1 / error_rate_for_op)  // 10x factor for CNOT
    ELSE: 
        error_rate_for_op = gate_error_rates.get(gate_name, 0.001) 
        node_weight = 1 * (1 / error_rate_for_op)     // 1x factor for others
    END IF
    ADD node_weight TO node_weights list
\end{lstlisting}
\textbf{3. HYPEREDGE CONSTRUCTION}
\begin{lstlisting}[basicstyle=\ttfamily\scriptsize]
Initialize empty hyperedges list and edge_weights list
3.1 Gate-Level Hyperedges (Multi-qubit interactions)
    For each gate_idx, gate in enumerate(circuit): 
        If gate acts on >1 qubit (e.g., CNOT):
            a. Create hyperedge containing only this gate_idx (0-based)
            b. Calculate edge weight:
                error_rate = gate_error_rates.get(gate.name, 0.05) 
                weight = 100 * num_qubits_in_gate * (1 / error_rate)
            c. Add to hyperedges and edge_weights lists
3.2 Temporal Chains (Qubit timelines)
    For each qubit_idx from 0 to num_qubits-1:
        a. Collect all gate_indices (0-based) acting on this qubit_idx into qubit_gate_indices
        b. If >1 gate exists in qubit_gate_indices:
            i. Create hyperedge containing all these gate_indices
            ii. Calculate temporal density:
                h_error_rate = gate_error_rates.get("HGate", 0.001)
                density = 100 * (len(qubit_gate_indices) // 2) * (1 / h_error_rate)
            iii. edge_weight = max(1, density)
            iv. Add to hyperedges and edge_weights lists
\end{lstlisting}
\textbf{4. DATA CLEANING}
\begin{lstlisting}[basicstyle=\ttfamily\scriptsize]
4.1 valid_hyperedges = []
4.2 valid_edge_weights = []
4.3 For he, w in zip(hyperedges, edge_weights):
4.4     If he is not empty:
4.5         valid_hyperedges.append(he)
4.6         valid_edge_weights.append(w)
\end{lstlisting}
\textbf{5. WEIGHT NORMALIZATION}
\begin{lstlisting}[basicstyle=\ttfamily\scriptsize]
Initialize empty normalized_weights list
If valid_edge_weights is not empty:
    5.1 Find max_weight in valid_edge_weights (ensure max_weight > 0)
    5.2 Normalize all weights:
        For each weight in valid_edge_weights:
            normalized_weight = int(round((weight * 1e6) / max_weight)) if max_weight > 0 else int(round(weight))
            Store normalized_weight in normalized_weights list
Else:
    normalized_weights = []
\end{lstlisting}
\textbf{6. FILE OUTPUT (hMETIS format)}
\begin{lstlisting}[basicstyle=\ttfamily\scriptsize]
6.1 Write header line:
    [len(valid_hyperedges)] [num_gates] 1  % Last '1' indicates (hyper)edge weights are present
6.2 Write each hyperedge:
    For hyperedge, norm_weight in zip(valid_hyperedges, normalized_weights):
        a. Convert node indices in hyperedge to 1-based numbering for hMETIS
        b. Write line: [norm_weight] [node1_1based node2_1based ...]
6.3 Write node weights:
    For weight in node_weights:
        Write line: [int(round(weight))] % Ensure integer node weights
\end{lstlisting}
\paragraph{Key Features Modeled} \begin{itemize} \item Gate Criticality: High-weight nodes for low-error (important) gates relative to their type. \item Multi-Qubit Penalty: Explicit hyperedges (containing the gate node) with weights discouraging cutting entangling gates from their context. \item Temporal Locality: Chains encourage grouping consecutive qubit operations with very high weights. \item Error Awareness: All weights inversely proportional to error rates or derived from them. \end{itemize}
\subsubsection{Complexity}  Overall complexity can be reasonably stated as $O(N_{\text{gates}} + N_{\text{qubits}})$ assuming efficient data structures.

\subsection{Partitioning (test\_kahypar\_partitioning)}
\label{subsec:partitioning_algo} 
\subsubsection{Input/Output}
The \texttt{test\_kahypar\_partitioning} function takes a hypergraph file path (\texttt{INPUT\_FILE}, default "circuit.hgr") generated from the circuit and an integer number of partitions (\texttt{NUM\_PARTITIONS}, default 2). It returns a list of \texttt{partition\_labels}, where each label corresponds to the partition assignment of a gate in the hypergraph.

\subsubsection{Dynamic Partition Count (\(k\))}
The number of partitions \(k\) is dynamically computed as \(k = \max(2, \min(N_{\text{ops}} / B, \lfloor\sqrt{N_{\text{qubits}}}\rfloor))\), where \(N_{\text{ops}}\) is the total number of operations (gates) in the circuit, \(N_{\text{qubits}}\) is the number of qubits, and \(B\) is the block size (typically set to a constant like 10, representing the maximum operations per partition for efficiency). This formula ensures a minimum of 2 partitions, caps the partition count based on operation distribution (\(N_{\text{ops}} / B\)), and limits it by the square root of qubits (\(\lfloor\sqrt{N_{\text{qubits}}}\rfloor\)) to balance computational load and hardware constraints in the Linear Chain topology, preventing excessive partitioning that could degrade fidelity.

\subsubsection{Mt-KaHyPar Configuration}
The Mt-KaHyPar tool is configured with the following parameters: \texttt{-o km1} for the k-way metric objective to optimize cut size, \texttt{-e 0.05} for a 5\% imbalance to allow flexibility in partition sizes, \texttt{--seed 42} for reproducibility, and \texttt{-t 16} for 16 threads to leverage multi-core processing. The \texttt{-m direct} initial partitioning mode and \texttt{--preset-type default} preset provide a robust starting point, while \texttt{--write-partition-file=true} ensures output is saved. These settings balance quality (km1, 5\% imbalance) and efficiency (threads, direct mode), suitable for NISQ hardware constraints and large hypergraphs from \texttt{Fidelipart}.

\subsubsection{Pseudocode} \label{algo:mtkh_algo} 
\paragraph{Input}
\begin{itemize}
    \item \texttt{INPUT\_FILE}: Hypergraph file path (default: "circuit.hgr")
    \item \texttt{NUM\_PARTITIONS}: Number of partitions (default: 2)
\end{itemize}
\paragraph{Output}
\begin{itemize}
    \item \texttt{partition\_labels}: List of partition assignments
\end{itemize}
\paragraph{Algorithm Steps}
\textbf{1. INITIALIZATION}
\begin{lstlisting}[basicstyle=\ttfamily\scriptsize]
1.1 Validate input parameters here if needed
1.2 Set paths and configuration:
    - mtkahypar_bin = "/path/to/mtkahypar/build/mt-kahypar/application/mtkahypar" 
    - partition_file = GENERATE_PARTITION_FILENAME(INPUT_FILE, NUM_PARTITIONS)
\end{lstlisting}
\textbf{2. COMMAND SETUP}
\begin{lstlisting}[basicstyle=\ttfamily\scriptsize]
2.1 Construct Mt-KaHyPar command:
    - cmd = [
        mtkahypar_bin,
        "-h", INPUT_FILE,                // Input hypergraph file
        "-k", STR(NUM_PARTITIONS),       // Number of partitions
        "-e", "0.05",                    // Epsilon (balance parameter)
        "-o", "km1",                     // Objective function (k-way metric)
        "-m", "direct",                  // Initial partitioning mode
        "--preset-type", "default",      // Configuration preset
        "--seed", "42",                  // Random seed for reproducibility
        "-t", "16",                      // Number of threads
        "--write-partition-file=true"    // Enable partition file output
    ]
\end{lstlisting}
\textbf{3. EXECUTION AND ERROR HANDLING}
\begin{lstlisting}[basicstyle=\ttfamily\scriptsize]
3.1 Execute partitioning command: EXECUTE_COMMAND(cmd)
3.2 If execution failed:
    - error_msg = GET_STDERR()
    - Throw error: "Mt-KaHyPar failed: " + error_msg
3.3 Verify output file creation:
    - If not FILE_EXISTS(partition_file):
        - Throw error: "Partition file not found: " + partition_file
\end{lstlisting}
\textbf{4. RESULT PROCESSING}
\begin{lstlisting}[basicstyle=\ttfamily\scriptsize]
4.1 Initialize empty partition_labels list
4.2 Open partition_file for reading as file_handle:
    - For each line in file_handle:
        - cleaned_line = TRIM_WHITESPACE(line)
        - If cleaned_line is not empty:
            - partition_labels.append(STRING_TO_INT(cleaned_line))
4.3 Optional debug output:
    - Print "Labels retrieved:"
    - Print partition_labels
\end{lstlisting}
\textbf{5. RETURN RESULTS}
\begin{lstlisting}[basicstyle=\ttfamily\scriptsize]
5.1 Return partition_labels
\end{lstlisting}
\paragraph{Key Features Modeled}
\begin{itemize}
    \item Dynamic Partitioning: Adapts \(k\) based on circuit size and hardware.
    \item Robust Execution: Handles errors and verifies output.
    \item Efficient Processing: Leverages multi-threading and reproducible seeds.
\end{itemize}

\subsection{Subcircuit Creation (\_create\_trimmed\_partitions )}
\label{subsec:subcircuit_trimming_algo}

\subsubsection{Trimming}
The \texttt{\_create\_trimmed\_partitions} function splits the original quantum circuit into subcircuits based on partition labels from \texttt{test\_kahypar\_partitioning} (Subsection \ref{subsec:partitioning_algo}). It takes the \texttt{circuit} (with all operations) and \texttt{labels} (one per operation, indicating partition assignment) as input, producing a list of trimmed subcircuits and corresponding global-to-physical qubit maps (local contiguous re-mapping). Initially, it validates that the number of labels matches the number of operations, ensuring consistency. For each unique partition ID (\texttt{part\_id}), it first collects all operations from the original circuit that were assigned this \texttt{part\_id}. This collection forms an \textit{initial subcircuit} conceptually still operating on the original number of qubits. It then identifies the set of \textit{active global qubits}—those original circuit qubit indices actually used by operations within this initial subcircuit—excluding idle qubits to optimize resource usage, particularly in constrained topologies like a Linear Chain. A global-to-physical qubit map (\texttt{physical\_map}) is created by sorting these active global qubits and assigning them consecutive local physical indices starting from 0. Finally, a \textit{trimmed subcircuit} is constructed using only these active qubits (now indexed locally via \texttt{physical\_map}), ensuring efficient execution on NISQ hardware by minimizing qubit overhead while preserving operation fidelity (e.g., avoiding unnecessary SWAP gates, where $\epsilon_{\text{SWAP}} \approx 0.03$, which itself implies multiple CNOTs with $\epsilon_{\text{CNOT}} = 0.05$).

\subsubsection{pseudocode}
\label{algo:trim_algo}
\paragraph{Input}
\begin{itemize}
    \item \texttt{circuit}: Original quantum circuit object (contains all operations and total qubit count)
    \item \texttt{labels}: List of partition assignments for each operation in \texttt{circuit.operations}
\end{itemize}
\paragraph{Output}
\begin{itemize}
    \item \texttt{final\_partitions}: List of trimmed subcircuits (each a new Circuit object)
    \item \texttt{final\_qubit\_maps}: List of dictionaries mapping global to local physical qubits for each partition
\end{itemize}
\paragraph{Algorithm Steps}

\textbf{1. INPUT VALIDATION}
\begin{lstlisting}[basicstyle=\ttfamily\scriptsize]
1.1 If length(labels) != length(circuit.operations):
    - Raise error: "Number of labels does not match number of operations in circuit"
\end{lstlisting}

\textbf{2. INITIALIZATION}
\begin{lstlisting}[basicstyle=\ttfamily\scriptsize]
2.1 Initialize final_partitions = [] 
2.2 Initialize final_qubit_maps = [] 
2.3 all_original_ops = circuit.operations
2.4 unique_part_ids = get_unique_sorted(labels)
\end{lstlisting}

\textbf{3. PARTITION PROCESSING LOOP}
\begin{lstlisting}[basicstyle=\ttfamily\scriptsize]
3.1 For each part_id in unique_part_ids:
    // 3.1.1 Collect operations for this partition_id
    - ops_for_this_partition = []
    - For op_idx, original_op in enumerate(all_original_ops):
        - If labels[op_idx] == part_id:
            - ops_for_this_partition.append(original_op)
    
    // 3.1.2 Identify active global qubits for these operations
    - active_global_qubits_for_part = empty set
    - For op_in_part in ops_for_this_partition:
        - Add all qubit indices from op_in_part.location to active_global_qubits_for_part
    
    // 3.1.3 Handle empty partition case
    - If not active_global_qubits_for_part:
        - Log warning: "Partition {part_id} is empty (no active qubits)."
        - Continue to next part_id // Or add empty circuit/map if policy dictates

    // 3.1.4 Create global-to-local physical mapping for this partition
    - sorted_active_global_indices = sort(list(active_global_qubits_for_part))
    - current_physical_map = new Dictionary()
    - For local_physical_idx, global_q_idx in enumerate(sorted_active_global_indices):
        - current_physical_map[global_q_idx] = local_physical_idx
        
    // 3.1.5 Construct the trimmed subcircuit
    - num_local_qubits = len(sorted_active_global_indices)
    - trimmed_subcircuit_for_part = new Circuit(num_local_qubits)
    - For original_op_for_part in ops_for_this_partition:
        - new_local_qubit_locations = []
        - For global_q_idx_in_op in original_op_for_part.location:
            - new_local_qubit_locations.append(current_physical_map[global_q_idx_in_op]) 
            % This assumes global_q_idx_in_op will always be in current_physical_map
            % due to active_global_qubits_for_part construction
        - remapped_op = create_operation(original_op_for_part.gate, 
                                         new_local_qubit_locations, 
                                         original_op_for_part.params)
        - trimmed_subcircuit_for_part.add_operation(remapped_op)
        
    // 3.1.6 Store results for this partition
    - final_partitions.append(trimmed_subcircuit_for_part)
    - final_qubit_maps.append(current_physical_map)
\end{lstlisting}

\textbf{4. RETURN RESULTS}
\begin{lstlisting}[basicstyle=\ttfamily\scriptsize]
4.1 Return (final_partitions, final_qubit_maps)
\end{lstlisting}
\paragraph{Key Features Modeled}
\begin{itemize}
    \item Optimized Resource Usage: Trims inactive qubits for NISQ efficiency.
    \item Qubit Remapping: Ensures compact local physical qubit indices for subcircuits.
    \item Robust Handling: Validates inputs and can log warnings for empty partitions.
\end{itemize}

\subsection{Subcircuit Merging ( merge\_partitions)}
\label{subsec:subcircuit_creation_merging_algo}
\subsubsection{Merging Rationale}
\label{subsec:merging_rationale_algo}
Merging is performed to reduce the number of cuts introduced after partitioning and trimming, which can lead to increased SWAP gate insertions (estimated logical realignments) and fidelity loss, particularly on quantum hardware with restricted connectivity like Linear Chain topologies (e.g., considering gate errors like \( \epsilon_{\text{SWAP}} \approx 0.03 \) which implies three CNOTs at $\epsilon_{\text{CNOT}} = 0.05$, and direct \( \epsilon_{\text{CNOT}} = 0.05 \)). Initial partitions may inadvertently split circuits across qubits with strong interactions, necessitating additional gates to bridge these gaps. By merging partitions that share a sufficient number of qubits (determined by a \texttt{threshold}), the algorithm aims to minimize these cuts. This consolidates operations into fewer, more cohesive subcircuits, thereby reducing the overhead associated with inter-partition communication and helping to preserve the integrity of the quantum state during execution.

\subsubsection{Merge Logic}
\label{subsec:merge_logic_algo}
The merge process, implemented in \texttt{\_merge\_partitions}, employs a \textbf{multi-pass, greedy approach}. It operates through a \texttt{while} loop that continues as long as merges were successfully performed in the previous pass. Within each pass, the algorithm iterates through all current partitions. For each partition (\texttt{p1}), it searches for the best possible merge partner (\texttt{p2}) among the subsequent partitions. The "best" partner is defined as the one that shares the highest number of original global qubits (\texttt{shared\_max}), provided this number meets or exceeds the specified \texttt{threshold}.

To manage the process, a set called \texttt{used\_indices\_current\_pass} tracks which partitions have been processed (either merged or kept) within the *current* pass. Another set, \texttt{merged\_pairs}, could track original identifiers of pairs that have been merged across *all* passes to prevent redundant merge attempts on the same original pair combination, though if indices are reassigned after each pass, its main effect might be within a pass or for specific re-combinations.

The \texttt{\_shared\_qubits} function calculates the overlap by taking the set of all global qubit keys from the respective qubit maps of the two partitions and finding their intersection.

If a suitable merge partner is found, \texttt{\_combine\_partitions} is called to create a new, larger circuit with an updated, unified qubit map. This new partition is added to a temporary list (\texttt{new\_partitions}), and the original partitions' indices are added to \texttt{used\_indices\_current\_pass}. Crucially, the \texttt{merged} flag is set to \texttt{True}, ensuring the \texttt{while} loop will execute at least one more time. If no suitable partner is found, the original partition is added to \texttt{new\_partitions} as is. At the end of each pass, the main \texttt{partitions} and \texttt{qubit\_maps} lists are updated with these new lists, and the process repeats until a full pass occurs with no merges.

\subsubsection{pseudocode}
\label{algo:merge_algo}
\paragraph{Input}
\begin{itemize}
    \item \texttt{partitions}: A list containing the individual quantum circuit partitions.
    \item \texttt{qubit\_maps}: A corresponding list of dictionaries mapping original qubit indices to local partition indices.
    \item \texttt{threshold}: An integer representing the minimum number of shared original qubits required for merging (default set by invocation, e.g., 3).
\end{itemize}
\paragraph{Output}
\begin{itemize}
    \item \texttt{partitions}: An updated list containing the final, potentially merged, quantum circuit partitions.
    \item \texttt{qubit\_maps}: The corresponding updated list of qubit maps.
\end{itemize}
\paragraph{Algorithm Steps}
\textbf{1. INITIALIZATION}
\begin{lstlisting}[firstnumber=1, basicstyle=\ttfamily\scriptsize]
1.1 Initialize merged = True  // Loop control flag
1.2 Initialize merged_original_pairs = empty_set() // Tracks stable identifiers of merged pairs
                                                 // e.g., frozenset of original partition IDs if available
                                                 // or frozenset of frozensets of map1.keys() & map2.keys()
\end{lstlisting}
\textbf{2. MULTI-PASS MERGE LOOP}
\begin{lstlisting}[basicstyle=\ttfamily\scriptsize]
2.1 While merged is True:
2.2     Set merged = False 
2.3     Initialize new_partitions_this_pass = []
2.4     Initialize new_qubit_maps_this_pass = []
2.5     Initialize used_indices_current_pass = empty_set()

2.6     For i from 0 to len(partitions)-1:
2.7         If i is in used_indices_current_pass: Continue

2.8         p1_current = partitions[i]
2.9         map1_current = qubit_maps[i]
        
2.10        shared_qubits_max_count = 0
2.11        best_merge_partner_idx = -1

2.12        For j from i+1 to len(partitions)-1:
2.13            If j is in used_indices_current_pass: Continue 
            
2.14            map2_candidate = qubit_maps[j]
            % Create a stable identifier for the pair (map1_current, map2_candidate)
            % E.g., pair_id = frozenset({frozenset(map1_current.keys()), frozenset(map2_candidate.keys())})
            % If pair_id in merged_original_pairs: Continue 
            
2.15            num_shared = len(calculate_shared_qubits(map1_current, map2_candidate))
            
2.16            If num_shared >= threshold AND num_shared > shared_qubits_max_count:
2.17                shared_qubits_max_count = num_shared
2.18                best_merge_partner_idx = j

2.19        If best_merge_partner_idx != -1:
2.20            p2_to_merge = partitions[best_merge_partner_idx]
2.21            map2_to_merge = qubit_maps[best_merge_partner_idx]
            
2.22            combined_circuit, combined_map = combine_partitions(p1_current, p2_to_merge, 
                                                                  map1_current, map2_to_merge)
            
2.23            Add combined_circuit to new_partitions_this_pass
2.24            Add combined_map to new_qubit_maps_this_pass
2.25            Add i to used_indices_current_pass
2.26            Add best_merge_partner_idx to used_indices_current_pass
            % Add stable identifier of (map1_current, map2_to_merge) to merged_original_pairs
2.27            Set merged = True 
2.28        Else:
2.29            Add p1_current to new_partitions_this_pass
2.30            Add map1_current to new_qubit_maps_this_pass
2.31            Add i to used_indices_current_pass 

2.32    Set partitions = new_partitions_this_pass
2.33    Set qubit_maps = new_qubit_maps_this_pass
\end{lstlisting}
\textbf{3. HELPER FUNCTIONS}
\begin{lstlisting}[basicstyle=\ttfamily\scriptsize]
3.1 function calculate_shared_qubits(map1, map2):
3.2     map1_global_keys = set(map1.keys())
3.3     map2_global_keys = set(map2.keys())
3.4     Return intersection(map1_global_keys, map2_global_keys)

3.5 function combine_partitions(p1, p2, map1, map2):
3.6     // 1. all_original_qubits = union(map1.keys(), map2.keys())
3.7     // 2. sorted_all_original_qubits = sort(list(all_original_qubits))
3.8     // 3. new_combined_map = {}
3.9     //    For new_local_idx, orig_q in enumerate(sorted_all_original_qubits):
3.10    //        new_combined_map[orig_q] = new_local_idx
3.11    // 4. combined_circuit = new Circuit(len(sorted_all_original_qubits))
3.12    // 5. For op in p1.operations:
3.13    //      remapped_loc = [new_combined_map[map1_inv[l_q]] for l_q in op.location] 
            // (Requires inverse map or direct access to original global indices for ops in p1)
            // Assuming ops in p1/p2 store original global qubit indices directly:
            //      remapped_loc = [new_combined_map[global_q] for global_q in op.global_location]
3.14    //      combined_circuit.add_operation(create_op(op.gate, remapped_loc, op.params))
3.15    // 6. Similarly for ops in p2.operations
3.16    Return combined_circuit, new_combined_map
\end{lstlisting}
\textbf{4. RETURN RESULTS}
\begin{lstlisting}[basicstyle=\ttfamily\scriptsize]
4.1 Return partitions, qubit_maps
\end{lstlisting}
\paragraph{Key Features Modeled} \begin{itemize}
    \item \textbf{Cut Reduction:} Merges partitions sharing significant qubit resources to minimize effective circuit cuts and reduce SWAP gate overhead.
    \item \textbf{Greedy Optimization:} In each pass, it selects the best merge partner based on the highest shared qubit count (above the threshold), aiming for efficient consolidation.
    \item \textbf{Multi-Pass Consolidation:} The \texttt{while} loop ensures the process repeats, allowing merges made in one pass to enable further beneficial merges in subsequent passes, until a stable configuration (no more merges) is reached.
    \item \textbf{Consistent Mapping:} Dynamically creates new qubit maps during merging, ensuring that all operations refer to a consistent, unified set of local qubit indices.
    \item \textbf{State Management:} Uses tracking sets (\texttt{used}, \texttt{merged\_pairs}) to manage the merging process efficiently within and across passes, according to the code's logic.
\end{itemize}

\subsection{Dependency Graph Construction (build\_dependency\_graph)}
\label{subsec:dependency_graph_algo}
The \texttt{build\_dependency\_graph} function constructs a directed acyclic graph (DAG) to model dependencies between circuit partitions after merging (Subsection \ref{algo:merge_algo}). It takes as input the list of merged partitions (\texttt{partitions}) and their corresponding global-to-physical qubit mappings (\texttt{qubit\_maps}). Each partition is represented as a node in the DAG, indexed from 0 to the number of partitions minus 1. The function evaluates all unique pairs of partitions. For each pair (partition $i$, partition $j$ where $i < j$), it checks if they share any original global qubits by examining the intersection of the keys in \texttt{qubit\_maps[i]} and \texttt{qubit\_maps[j]}. If such shared qubits exist, a directed edge is added from node $i$ to node $j$, establishing a dependency. This ensures that dependencies account for qubit interactions critical in the Linear Chain topology. Finally, an optional \texttt{print\_dependencies} function can provide diagnostic output, detailing each edge and the shared qubits, aiding in understanding partition interactions and their impact on fidelity (e.g., by informing strategies to minimize SWAP gates, with $\epsilon_{\text{SWAP}} = 0.03$).
\subsubsection{Pseudocode}
\label{algo:DAG_algo}
\paragraph{Input}
\begin{itemize}
    \item \texttt{partitions}: List of merged subcircuits
    \item \texttt{qubit\_maps}: List of global-to-physical qubit mappings for each partition
\end{itemize}
\paragraph{Output}
\begin{itemize}
    \item \texttt{DAG}: Directed acyclic graph representing partition dependencies
\end{itemize}
\paragraph{Algorithm Steps}
\textbf{1. INITIALIZATION}
\begin{lstlisting}[basicstyle=\ttfamily\scriptsize]
1.1 Initialize an empty directed graph (DAG)
1.2 For p_idx from 0 to len(partitions)-1:
1.3     Add node p_idx to DAG
\end{lstlisting}
\textbf{2. DEPENDENCY DETECTION}
\begin{lstlisting}[basicstyle=\ttfamily\scriptsize]
2.1 For i from 0 to len(partitions) - 1:
    - For j from i + 1 to len(partitions) - 1: // Ensures i < j and unique pairs
        - global_qubits_in_map_i = set(qubit_maps[i].keys())
        - global_qubits_in_map_j = set(qubit_maps[j].keys())
        - shared_global_qubits = intersection(global_qubits_in_map_i, global_qubits_in_map_j)
            
        - If shared_global_qubits is not empty:
            - Add directed edge from node i to node j in DAG
            - Optionally, store shared_global_qubits as an attribute of the edge (i,j)
\end{lstlisting}
\textbf{3. DIAGNOSTIC OUTPUT (Optional)}
\begin{lstlisting}[basicstyle=\ttfamily\scriptsize]
3.1 For each edge (u, v) in DAG.edges:
3.2     Print "Partition {u} -> Partition {v} | Shared qubits: {edge_attribute_shared_qubits}"
3.3 Print "Total dependencies: {DAG.num_edges}" 
\end{lstlisting}
\textbf{4. RETURN RESULTS}
\begin{lstlisting}[basicstyle=\ttfamily\scriptsize]
4.1 Return DAG
\end{lstlisting}
\paragraph{Key Features Modeled}
\begin{itemize}
    \item Dependency Tracking: Captures qubit-sharing relationships (based on original global indices) between partitions.
    \item Scheduling Support: Directed edges guide execution order.
    \item Diagnostic Clarity: Can print detailed dependency information for analysis.
\end{itemize}

\paragraph{Algorithm Steps}

\textbf{1. INITIALIZATION}
\begin{lstlisting}[basicstyle=\ttfamily\scriptsize]
1.1 Initialize an empty directed graph (DAG)
1.2 For p_idx from 0 to len(partitions)-1:
1.3     Add node p_idx to DAG
\end{lstlisting}

\textbf{2. DEPENDENCY DETECTION}
\begin{lstlisting}[basicstyle=\ttfamily\scriptsize]
2.1 For i from 0 to len(partitions) - 1:
    - For j from i + 1 to len(partitions) - 1: // Ensures i < j and unique pairs
        // Check if partitions share original global qubits based on their maps
        - map_i_global_qubits = set(qubit_maps[i].keys())
        - map_j_global_qubits = set(qubit_maps[j].keys())
        - shared_global_qubits_for_edge_ij = intersection(map_i_global_qubits, map_j_global_qubits)
            
        - If shared_global_qubits_for_edge_ij is not empty:
            - Add directed edge from node i to node j in DAG
            - Optionally, store shared_global_qubits_for_edge_ij as an attribute of the edge (i,j)
\end{lstlisting}

\textbf{3. DIAGNOSTIC OUTPUT (Optional)}
\begin{lstlisting}[basicstyle=\ttfamily\scriptsize]
3.1 For each edge (u, v) in DAG.edges:
3.2     % Retrieve shared_global_qubits_for_edge_uv if stored as edge attribute
3.3     Print "Partition {u} -> Partition {v} | Shared qubits: {shared_global_qubits_for_edge_uv}"
3.4 Print "Total dependencies: {DAG.num_edges}" 
\end{lstlisting}

\textbf{4. RETURN RESULTS}
\begin{lstlisting}[basicstyle=\ttfamily\scriptsize]
4.1 Return DAG
\end{lstlisting}

\paragraph{Key Features Modeled}
\begin{itemize}
    \item Dependency Tracking: Captures qubit-sharing relationships (based on original global indices) between partitions.
    \item Scheduling Support: Directed edges guide execution order.
    \item Diagnostic Clarity: Can print detailed dependency information for analysis.
\end{itemize}

\subsection{Comparison Methodology}
\label{subsec:comparison_methodology_algo_main} 

To rigorously evaluate the performance of \texttt{Fidelipart}, we conducted a comparative analysis against BQSKit's \texttt{QuickPartitioner}. Both partitioning strategies were applied to the same set of quantum circuits and evaluated under identical hardware assumptions using a consistent suite of performance metrics. This ensures a fair and direct comparison of their effectiveness in managing circuit complexity and mitigating errors.

The evaluation assumes a \textbf{linear qubit topology} for the underlying hardware. Following the initial partitioning (by either \texttt{Fidelipart} or \texttt{QuickPartitioner}), which yields subcircuits defined by gates acting on original global qubit indices, a consistent \textbf{local contiguous re-mapping} strategy is applied to each resulting subcircuit. In this re-mapping, active global qubits within each partition are identified, sorted, and then assigned new, contiguous local physical indices starting from 0. This normalization is critical for standardized subcircuit representation and ensures a fair basis for calculating inter-partition communication costs, such as SWAP gate overhead.

\subsubsection{Baseline Setup: \texttt{QuickPartitioner}}
\label{subsubsec:baseline_quickpartitioner_algo_main} 

The \texttt{QuickPartitioner} from BQSKit\cite{bqskit2021} (version 1.2.0) was selected as the baseline, representing a standard heuristic partitioning approach within the BQSKit framework. Our setup for \texttt{QuickPartitioner} was as follows:

\begin{itemize}
    \item \textbf{Invocation:} \texttt{QuickPartitioner} was initialized with a configurable \texttt{block\_size} parameter (e.g., 4, 6 and 8 based on the circuit size) to guide its logical gate grouping.
    \item \textbf{Hardware Context:} The partitioning was performed in the context of a \texttt{HardwareGraph} instance, configured with a linear topology and the total number of qubits corresponding to the input circuit. This same hardware model was consistently used for both \texttt{QuickPartitioner} and \texttt{Fidelipart}.
    \item \textbf{Qubit Mapping and Subcircuit Finalization:} The \texttt{partition\_with\_quick} function processes the output of BQSKit's \texttt{QuickPartitioner}\cite{bqskit2021} to yield an initial list of subcircuits where gates operate on their original global qubit indices. Subsequently, the \texttt{derive\_and\_remap\_for\_quick} function applies the "local contiguous re-mapping" strategy to each of these subcircuits. This involves the following steps for each subcircuit:
    \begin{enumerate}
        \item The set of actively utilized global qubit indices within the subcircuit is identified.
        \item These active global indices are then sorted.
        \item A new \texttt{\{global\_qubit: local\_physical\_qubit\}} map is created, assigning contiguous local physical indices (starting from 0) to these sorted global qubits.
        \item Finally, a new, remapped subcircuit is constructed where all gate operations utilize these new local physical indices.
    \end{enumerate}
    This two-step process (initial partitioning followed by the re-mapping) ensures that the partitions derived from \texttt{QuickPartitioner} are structured identically to those produced by \texttt{Fidelipart}'s internal \texttt{\_create\_trimmed\_partitions} method before metric calculation. This standardization of the local physical view is paramount for a fair and direct comparison of the two partitioning strategies.
\end{itemize}

\subsubsection{Evaluation Workflow}
\label{subsubsec:evaluation_workflow_algo_main} 

The entire comparative evaluation was orchestrated by the \texttt{compare\_partitions\_physical} asynchronous function (though `await` is removed from pseudocode for generality, the original implementation may be asynchronous). For each input quantum circuit, this function executes the following sequential workflow:

\begin{enumerate}
    \item Initializes the \texttt{HardwareGraph} instance based on the specified topology (linear) and the total number of qubits in the input circuit.
    
    \item Processes the circuit using the \texttt{QuickPartitioner} path:
    \begin{itemize}
        \item Invokes the \texttt{partition\_with\_quick} function to obtain an initial list of subcircuits, where gates operate on their original global qubit indices.
        \item Calls the \texttt{derive\_and\_remap\_for\_quick} function on these global-indexed subcircuits. This function applies the local contiguous re-mapping strategy to generate the final local-indexed subcircuits and their corresponding \texttt{\{global\_qubit: local\_physical\_qubit\}} maps.
    \end{itemize}
    
    \item Processes the circuit using the \texttt{Fidelipart} path:
    \begin{itemize}
        \item Invokes the \texttt{partition\_with\_hypergraph} function. This function encapsulates the \texttt{Fidelipart} pipeline, which internally uses \texttt{EnhancedHypergraphPartitionPass}. The pass sequentially calls \texttt{circuit\_to\_hypergraph} for hypergraph construction, \texttt{test\_kahypar\_partitioning} for partitioning via Mt-KaHyPar, \texttt{\_create\_trimmed\_partitions} for subcircuit generation with local contiguous re-mapping, and includes partition merging and dependency graph construction steps. This path directly yields local-indexed subcircuits and their \texttt{\{global\_qubit: local\_physical\_qubit\}} maps.
    \end{itemize}
    
    \item For both sets of resulting local-indexed subcircuits and their associated maps (from \texttt{QuickPartitioner} and \texttt{Fidelipart}), it calls the \texttt{\_calculate\_physical\_partition\_metrics} function to compute the defined performance indicators, such as cut qubits, SWAP overhead, and fidelity.
    
    \item The \texttt{\_validate\_gate\_counts} function is then called. This function ensures that the core gate operations (excluding any SWAP gates introduced for inter-partition communication) are conserved in the combined set of partitioned subcircuits compared to the original input circuit, serving as a sanity check for both partitioning methods.
    
    \item Finally, the \texttt{\_format\_comparison\_results} function collates and presents the calculated performance metrics for both \texttt{QuickPartitioner} and \texttt{Fidelipart}, facilitating a direct comparison.
\end{enumerate}

\subsubsection{Performance Metrics}
\label{subsubsec:performance_metrics_unified}

The performance metrics define the quantitative framework for our comparative analysis. Each metric captures distinct aspects of partitioning quality, such as communication overhead, estimated circuit fidelity, and computational demands. Below, we detail their operational definitions, significance, and measurement protocols:

\begin{description}[leftmargin=*, font=\normalfont, labelwidth=3.5cm, labelsep=0.5em, style=nextline, itemsep=0.5em] 
    \item[\faLink \quad Cut Qubit Count]  
    \textbf{Definition:} The number of unique global qubit indices that are present in the \texttt{\{global\_qubit: local\_physical\_qubit\}} map of more than one partition. \\
    \textbf{Significance:} Indicates the extent of inter-partition qubit sharing and thus the communication requirements. \textbf{(Lower is better)} \\
    \textbf{Measurement:} Identified by the \texttt{get\_physical\_cut\_qubits} function (which conceptually analyzes global qubit indices) by examining the \texttt{qubit\_maps} for global qubits common to multiple partitions.

    \item[\faExchange \quad SWAP Gates (Estimated Logical Realignments)] 
    \textbf{Definition:} The estimated number of SWAP operations required for the logical realignment of shared global qubits across partition boundaries, based on differing local physical index assignments in their respective contiguous re-mappings. \\ 
    \textbf{Significance:} Measures the direct communication cost arising from logical conflicts at partition interfaces and represents a significant source of circuit error. \textbf{(Lower is better)} \\
    \textbf{Measurement Details:} This estimation, performed by the \texttt{compute\_physical\_swaps\_between\_partitions} function, incorporates the following:
        \begin{itemize}[noitemsep, topsep=0pt, partopsep=0pt, leftmargin=*, itemsep=0.2em]
            \item A cost of 1 SWAP is incurred for each shared global qubit that is mapped to different local physical indices in two partitions.
            \item A teleportation heuristic is applied: if a specific global qubit incurs more than 3 such SWAP misalignments across different cuts, there is a 60\% probability that subsequent SWAP requirements for that qubit are waived. This heuristic (with \texttt{random.seed(42)} for reproducibility) models a potential shift to quantum teleportation for extensive state transfers.
            \item The total SWAP cost for communication between a pair of partitions is attributed to the partition with the lower index in the pair.
        \end{itemize}
    This metric quantifies logical misalignments rather than the full, distance-dependent routing cost on the linear chain, thus serving as a lower bound on the actual physical SWAP overhead. A more detailed routing analysis is beyond the scope of this partitioning study.

    \item[\faCheckCircle \quad Estimated Fidelity]  
    \textbf{Definition:} Circuit fidelity ($F$) is estimated using a product-of-errors model. For each partition $p$, its fidelity $F_p$ is given by:
    \begin{equation}
    \label{eq:fidelity} 
        F_p = (1 - \epsilon_H)^{N_H^{(p)}} \times (1 - \epsilon_{\text{CNOT}})^{(N_{\text{CNOT}}^{(p)} + 3 N_{\text{SWAP}}^{(p)})}
    \end{equation}
    Here, $N_H^{(p)}$ is the count of Hadamard gates, $N_{\text{CNOT}}^{(p)}$ is the count of CNOT gates, and $N_{\text{SWAP}}^{(p)}$ is the number of SWAP gates attributed to partition $p$ for inter-partition communication (each SWAP is modeled as three CNOT operations for its error contribution). The total circuit fidelity is $F = \prod_p F_p$. \\
    \textbf{Model Assumptions:} The assumed single-qubit gate error rate is $\epsilon_H = 0.001$, and the two-qubit gate error rate is $\epsilon_{\text{CNOT}} = 0.05$. While Equation~\ref{eq:fidelity} explicitly shows terms for H, CNOT, and SWAP gates, the \texttt{compute\_fidelity} function accounts for other gate types (e.g., CCX) by considering their decomposition into this base set for error accumulation, or by applying a default single-qubit error rate for other elementary single-qubit operations. \\
    \textbf{Significance:} Quantifies the overall noise resilience of the partitioned circuit and the likelihood of successful computation. \textbf{(Higher is better)} \\
    \textbf{Measurement:} Calculated using the \texttt{compute\_fidelity} function based on the defined model and assumed error rates.

    \item[\faClockO \quad Maximum Depth]  
    \textbf{Definition:} The longest path of causally dependent gates (critical path) within any single partition. \\
    \textbf{Significance:} Relates to the potential parallel execution time of the subcircuits and susceptibility to decoherence for the longest-running segment. \textbf{(Lower is better)} \\
    \textbf{Measurement:} Obtained via the \texttt{partition.depth} attribute provided by BQSKit for each subcircuit; the maximum among these is reported.

    \item[\faHourglass \quad Partitioning Time]  
    \textbf{Definition:} The total wall-clock time taken to execute the entire partitioning and local re-mapping process for each method. \\
    \textbf{Significance:} Measures the computational efficiency of the partitioning algorithm itself. \textbf{(Lower is better)} \\
    \textbf{Measurement:} Recorded using \texttt{time.time()} calls bracketing the execution of the main partitioning functions (i.e., \texttt{partition\_with\_quick} combined with \texttt{derive\_and\_remap\_for\_quick} for the baseline, and \texttt{partition\_with\_hypergraph} for \texttt{Fidelipart}).
\end{description}

\section{Results}
\label{sec:results_main} 
This section presents the empirical results from the comparative analysis of \texttt{Fidelipart} and \texttt{QuickPartitioner}. The evaluation was performed using the methodology detailed in Section~\ref{subsec:comparison_methodology_algo_main}, with a consistent CNOT error rate $\epsilon_{\text{CNOT}} = 0.05$ used for both hypergraph weighting and final fidelity estimations. We analyze performance across benchmark circuits of varying sizes and complexities.

\subsection{Benchmark Circuits and Experimental Parameters}
\label{subsec:benchmarks_experimental_params_results_main} 

We selected a suite of benchmark circuits with varying sizes and structural complexities to evaluate the partitioning algorithms. For each circuit, the specified \texttt{block\_size} was used both as input for \texttt{QuickPartitioner} and in the formula to determine the target number of partitions, $k$, for \texttt{Fidelipart} ($k = \max(2, \min(\lfloor N_{ops} / \text{block\_size} \rfloor, \lfloor \sqrt{N_{qubits}} \rfloor))$). The BQSKit version used for this study was 1.2.0.

\begin{itemize}
    \item \textbf{Small-Scale Circuit (``Circuit S''):} 
    A 6-qubit, 22-gate circuit constructed with layered Hadamard gates and CNOT gates whose locations were determined using \texttt{random.sample(range(6), 2)} with a fixed seed (\texttt{random.seed(42)}) for deterministic generation, creating varied entanglement patterns. 
    For this circuit, a \texttt{block\_size} of 4 was used, resulting in \texttt{Fidelipart} targeting $k=2$ partitions.

    \item \textbf{Intermediate-Scale Circuit (``Circuit M''):} 
    A deterministically generated 10-qubit, 55-gate circuit. Its structure involves layered Hadamards, a linear chain of CNOTs for local entanglement, and specifically chosen wider-stride CNOTs to establish crossed entanglement patterns.
    For this circuit, a \texttt{block\_size} of 6 was used, resulting in \texttt{Fidelipart} targeting $k=3$ partitions.

    \item \textbf{Large-Scale Entangled Circuit (``Circuit L''):} 
    A 24-qubit, 88-gate highly entangled circuit, incorporating layered CNOTs and CCX gates, designed to rigorously test the scalability and efficacy of the partitioning algorithms. 
    For this circuit, a \texttt{block\_size} of 8 was used, resulting in \texttt{Fidelipart} targeting $k=4$ partitions.
\end{itemize}

For all evaluations, the underlying hardware model assumed a linear qubit topology. Full details of the comparison methodology, including the consistent local contiguous re-mapping applied to the outputs of both methods before metric calculation, are provided in Section~\ref{subsec:comparison_methodology_algo_main}.

\subsection{Illustrative Walkthrough of \texttt{Fidelipart}'s Stages}
\label{subsec:walkthrough_results_main} 

To provide insight into the operational flow of \texttt{Fidelipart}, we present an illustrative walkthrough using the Small-Scale Circuit (``Circuit S''), a 6-qubit, 22-gate circuit, with \texttt{Fidelipart} configured via a \texttt{block\_size} of 6, targeting $k=2$ partitions.

\subsubsection{ Input Circuit Snippet and Fidelity-Aware Hypergraph Representation}
\label{para:walkthrough_input_hypergraph_results_main} 

The input ``Circuit S'' (6-qubits, 22-gates) is defined as follows:
\begin{lstlisting}[caption={Gate sequence for ``Circuit S''}, basicstyle=\ttfamily\tiny]
Circuit structure:
Gate  0: HGate on (0,)
Gate  1: HGate on (3,)
Gate  2: CNOTGate on (5, 0)
Gate  3: HGate on (0,)
Gate  4: CNOTGate on (1, 5)
Gate  5: CNOTGate on (0, 2)
Gate  6: HGate on (1,)
Gate  7: CNOTGate on (5, 4)
Gate  8: HGate on (0,)
Gate  9: HGate on (2,)
Gate 10: CNOTGate on (1, 0)
Gate 11: HGate on (2,)
Gate 12: CNOTGate on (0, 4)
Gate 13: HGate on (2,)
Gate 14: CNOTGate on (3, 0)
Gate 15: HGate on (4,)
Gate 16: CNOTGate on (0, 5)
Gate 17: HGate on (4,)
Gate 18: CNOTGate on (1, 5)
Gate 19: HGate on (4,)
Gate 20: HGate on (4,)
Gate 21: CNOTGate on (4, 5)
\end{lstlisting}

This circuit is first transformed into a fidelity-aware hypergraph by the \texttt{circuit\_to\_hypergraph} process (Algorithm~\ref{algo:circuit_to_hypergrap_algo} ). This transformation involves creating nodes representing each of the 22 gates and two primary types of weighted hyperedges designed to capture quantum-specific dependencies and error characteristics (using $\epsilon_{\text{CNOT}}=0.05$ and $\epsilon_H=0.001$):

\begin{itemize}
    \item \textbf{Gate-Level Hyperedges (Multi-Qubit Interactions):} Each multi-qubit gate (primarily \texttt{CNOTGate}s in this example) is represented as a distinct hyperedge that contains only the index of that gate itself. For instance, \texttt{CNOTGate} at index 2 (gates are 0-indexed internally, but 1-indexed in the hMETIS file, e.g., gate 3 in the file) would form such a hyperedge. The weight of this type of hyperedge is calculated as $w_g = C_g \times N_q \times (1/\epsilon_g)$, where $C_g$ is a scaling factor (e.g., 100), $N_q$ is the number of qubits the gate acts upon, and $\epsilon_g$ is the gate's error rate (e.g., 0.05 for CNOTs). This assigns a high penalty to cutting through error-prone multi-qubit operations, guiding the partitioner to keep them intact within a single partition.

    \item \textbf{Temporal Chain Hyperedges (Qubit Lifetimes):} For each qubit, a hyperedge is created that connects all gate indices acting sequentially on that particular qubit. These hyperedges capture the "lifetime" or temporal evolution of a qubit's state. Their weights are determined by a temporal density metric, $w_t = \max(1, C_t \times (\lfloor N_{ops\_on\_q} / 2 \rfloor) \times (1/\epsilon_{ref}))$, where $C_t$ is a scaling factor (e.g., 100), $N_{ops\_on\_q}$ is the number of operations on the qubit, and $\epsilon_{ref}$ is a reference error rate (e.g., 0.001 for an \texttt{HGate}). This weighting penalizes the splitting of long, coherent sequences of operations on individual qubits, promoting data locality.
\end{itemize}

Node weights for each gate are also calculated, primarily based on the inverse of their respective error rates (e.g., $w_n \approx C_n \times (1/\epsilon_n)$), with a higher scaling factor $C_n$ for critical or more error-prone gates like \texttt{CNOTGate} (e.g., $10 \times$) compared to others (e.g., $1 \times$). All hyperedge weights are subsequently normalized (e.g., scaled to a maximum of $10^6$) for compatibility with the hypergraph partitioner.

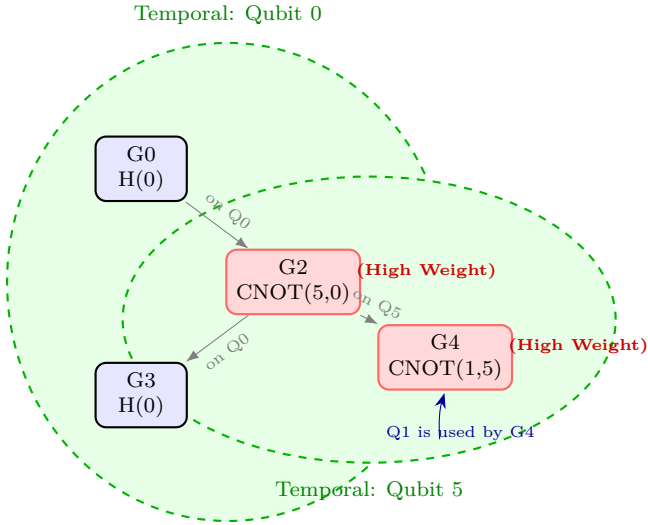
\begin{figure}[h!]
\centering
\begin{tikzpicture}[
    gate/.style={rectangle, draw, rounded corners, fill=blue!10, thick,
                 minimum height=0.8cm, minimum width=1.2cm, font=\scriptsize, align=center},
    cnot_gate/.style={gate, fill=red!15, draw=red!60},
    temporal_hyperedge/.style={ellipse, draw=green!70!black, fill=green!10, thick, dashed,
                                minimum width=1.5cm, minimum height=1cm}, 
    node_label/.style={font=\tiny\bfseries, text=black},
    hyperedge_label/.style={font=\scriptsize, text=green!50!black}
]
\node[gate] (g0) at (0,4) {G0\\H(0)};
\node[cnot_gate] (g2) at (2,2.5) {G2\\CNOT(5,0)};
\node[gate] (g3) at (0,1) {G3\\H(0)};
\node[cnot_gate] (g4) at (4,1.5) {G4\\CNOT(1,5)};
\node[gate, draw=none, fill=none] (q1_anchor_for_g4) at (4,0) {}; 
\begin{pgfonlayer}{background}
    \node[temporal_hyperedge, fit=(g0) (g2) (g3), inner sep=0.3cm] (th_q0) {};
    \node[hyperedge_label, above=0.1cm of th_q0.north] {Temporal: Qubit 0};
    \node[temporal_hyperedge, fit=(g2) (g4), inner sep=0.4cm] (th_q5) {};
    \node[hyperedge_label, below=0.1cm of th_q5.south] {Temporal: Qubit 5};
\end{pgfonlayer}
\node[node_label, text=red!80!black, below right=0.05cm and -0.2cm of g2.north east] {(High Weight)};
\node[node_label, text=red!80!black, below right=0.05cm and -0.2cm of g4.north east] {(High Weight)};
\draw[gray, thin, -Latex] (g0) -- (g2) node[midway, above, sloped, font=\tiny] {on Q0};
\draw[gray, thin, -Latex] (g2) -- (g3) node[midway, below, sloped, font=\tiny] {on Q0};
\draw[gray, thin, -Latex] (g2) -- (g4) node[midway, above, sloped, font=\tiny] {on Q5};
\draw (4.2,0.5) node[font=\tiny, text=blue!60!black] {Q1 is used by G4};
\draw[-{Stealth[length=2mm, width=1.5mm]}, blue!60!black, shorten >=1pt] (q1_anchor_for_g4) to [bend left=10] (g4.south);
\end{tikzpicture}
\caption{Conceptual visualization of a fidelity-aware hypergraph segment for ``Circuit S''. Gate nodes are shown as rectangles (blue for single-qubit, red for multi-qubit). Multi-qubit gates G2 and G4 inherently represent high-weight gate-level hyperedges, discouraging cuts through them. Dashed ellipses depict temporal chain hyperedges, grouping gates that act on the same qubit over time (e.g., Qubit 0 involving G0, G2, G3; Qubit 5 involving G2, G4). These hyperedges are weighted according to error rates ($\epsilon_{\text{CNOT}}=0.05, \epsilon_H=0.001$) and structural properties (e.g., temporal density), guiding the partitioner towards fidelity-preserving solutions.}
\label{fig:hypergraph_conceptual_results_main}
\end{figure}

The resulting hypergraph for ``Circuit S'', in the hMETIS format (weights reflect $\epsilon_{\text{CNOT}}=0.05$), is as follows:
\begin{lstlisting}[caption={hMETIS data for ``Circuit S'' hypergraph (weights reflect $\epsilon_{\text{CNOT}}=0.05$)}, basicstyle=\ttfamily\tiny]
16 22 1
4000 3
4000 5
4000 6
4000 8
4000 11
4000 13
4000 15
4000 17
4000 19
4000 22
400000 1 3 4 6 9 11 13 15 17
200000 5 7 11 19
200000 6 10 12 14
100000 2 15
300000 8 13 16 18 20 21 22
300000 3 5 8 17 19 22
1000
1000
200
1000
200
200
1000
200
1000
1000
200
1000
200
1000
200
1000
200
1000
200
1000
1000
200
\end{lstlisting}
Interpreting this hMETIS file:
\begin{itemize}
    \item The first line, \texttt{16 22 1}, indicates 16 hyperedges, 22 nodes (gates, indexed 1 to 22 in the file, corresponding to 0-indexed gates 0-21), and the '1' signifies that the hyperedges are weighted (node weights are listed separately at the end).
    \item Lines 2-17 (e.g., \texttt{4000 3}) define the hyperedges. The first number is the (normalized, in practice, though shown here as pre-normalization for clarity) hyperedge weight, followed by the 1-indexed gate numbers that belong to that hyperedge. For example, \texttt{4000 3} is a gate-level hyperedge for gate at original index 2 (CNOT), and \texttt{400000 1 3 ... 17} is a temporal chain hyperedge for qubit 0.
    \item Lines 18-39 (e.g., \texttt{1000}) list the individual node (gate) weights for gates 1 through 22, respectively. A weight of \texttt{1000} corresponds to an \texttt{HGate} ($1/0.001$), and \texttt{200} to a \texttt{CNOTGate} ($10 \times (1/0.05)$).
\end{itemize}
This weighted structure is designed so that a hypergraph partitioner, aiming to minimize the total weight of cut hyperedges (the ``cut''), will be naturally discouraged from splitting critical multi-qubit operations or disrupting coherent sequences of operations on the same qubit. By assigning higher weights (costs) to cutting through error-prone gates or dense temporal chains, \texttt{Fidelipart} biases the partitioning towards solutions that are more likely to preserve the quantum state's integrity and enhance overall circuit fidelity. A conceptual visualization of this transformation for a small segment of ``Circuit S'' is shown in Figure~\ref{fig:hypergraph_conceptual_results_main}.

\subsubsection{Partitioning and Trimming}
\label{para:walkthrough_partitioning_trimming_results_main}

The generated hypergraph is then partitioned by Mt-KaHyPar, as detailed in the \texttt{test\_kahypar\_partitioning} process (Algorithm~\ref{algo:mtkh_algo}), targeting $k=2$ partitions. For ``Circuit S'', this yields the following gate-to-partition assignments (labels):
\begin{lstlisting}[basicstyle=\ttfamily\tiny]
Labels retrieved:
[0, 0, 1, 0, 1, 0, 0, 1, 0, 0, 0, 0, 1, 
0, 0, 1, 1, 1, 1, 1, 1, 1]
\end{lstlisting}
Here, gates $0, 1, 3, 5, \dots$ (0-indexed from original circuit) are assigned to partition 0, and gates $2, 4, 7, \dots$ are assigned to partition 1.

These labels are then used by the \texttt{\_create\_trimmed\_partitions} process (Algorithm~\ref{algo:trim_algo}). This step first collects all gates from the original circuit belonging to each partition ID. Then, for each collection of gates, it identifies the set of unique \textit{global} qubit indices actively used. Finally, it creates a \texttt{\{global\_qubit: local\_physical\_qubit\}} map by sorting these active global qubits and assigning them contiguous local indices starting from 0. This also defines the size of the trimmed local circuit that would be constructed.

For ``Circuit S'', the state after this initial partitioning and mapping (and before any merge operations) is as follows, directly from our trace:

\textbf{Partition 0 (Initial):}
\begin{itemize}
    \item Gates (specified by their original global qubit indices):
    \begin{lstlisting}[basicstyle=\ttfamily\tiny]
  - HGate on (0,)
  - HGate on (1,)
  - HGate on (3,)
  - HGate on (0,)
  - CNOTGate on (0, 2)
  - HGate on (0,)
  - HGate on (2,)
  - CNOTGate on (1, 0)
  - HGate on (2,)
  - HGate on (2,)
  - CNOTGate on (3, 0)
    \end{lstlisting}
    \item Qubit Map (\texttt{\{global: local\}}): \texttt{\{0: 0, 1: 1, 2: 2, 3: 3\}}
    (Indicating active global qubits \{0, 1, 2, 3\} for this partition)
\end{itemize}

\textbf{Partition 1 (Initial):}
\begin{itemize}
    \item Gates (specified by their original global qubit indices, matching the order of assignment to Partition 1 based on the original circuit):
    \begin{lstlisting}[basicstyle=\ttfamily\tiny]
  - CNOTGate on (5, 0)  % Original Gate 2
  - CNOTGate on (1, 5)  % Original Gate 4
  - CNOTGate on (5, 4)  % Original Gate 7
  - CNOTGate on (0, 4)  % Original Gate 12
  - HGate on (4,)       % Original Gate 15
  - CNOTGate on (0, 5)  % Original Gate 16
  - HGate on (4,)       % Original Gate 17
  - CNOTGate on (1, 5)  % Original Gate 18
  - HGate on (4,)       % Original Gate 19
  - HGate on (4,)       % Original Gate 20
  - CNOTGate on (4, 5)  % Original Gate 21
    \end{lstlisting}
    \item Qubit Map (\texttt{\{global: local\}}): \texttt{\{0: 0, 1: 1, 4: 2, 5: 3\}}
    (Indicating active global qubits \{0, 1, 4, 5\} are used by the gates in this partition)
\end{itemize}

This step effectively prepares the subcircuits for potential execution on smaller devices or for further analysis, with each subcircuit having a well-defined local qubit space and a clear mapping back to the original circuit's global qubit indices.

\subsubsection{Merging Process}
\label{para:walkthrough_merging_results_main}

\texttt{Fidelipart} includes an optional iterative merging step (\texttt{\_merge\_partitions}, Algorithm~\ref{algo:merge_algo}) to combine partitions that share a significant number of qubits above a defined threshold. For this illustrative run with ``Circuit S'', a merge was attempted between Partition 0 and Partition 1:
\begin{lstlisting}[basicstyle=\ttfamily\tiny]
Before merging partitions 0 and 1:
Partition 0 gates: ['HGate@(0,)', 
'HGate@(1,)', 'HGate@(3,)', 'HGate@(0,)',
'CNOTGate@(0, 2)', 'HGate@(0,)', 
'HGate@(2,)', 'CNOTGate@(1, 0)', 
'HGate@(2,)', 'HGate@(2,)', 
'CNOTGate@(3, 0)']
Partition 0 map: {0: 0, 1: 1, 2: 2, 3: 3}
Partition 1 gates: ['CNOTGate@(5,0)',
'CNOTGate@(1,5)', 'CNOTGate@(5,4)', 
'CNOTGate@(0,4)', 'CNOTGate@(0,5)', 
'HGate@(4,)', 'CNOTGate@(1,5)', 
'HGate@(4,)', 'HGate@(4,)', 'HGate@(4,)',
'CNOTGate@(4,5)']
Partition 1 map: {0: 0, 1: 1, 4: 2, 5: 3} % Global indices {0,1,4,5} are used by these ops.

After merging partitions 0 and 1:
Merged partition gates: ['HGate@(0,)', 
'HGate@(1,)', 'HGate@(3,)', 'HGate@(0,)', 
'CNOTGate@(0, 2)', 'HGate@(0,)', 
'HGate@(2,)', 'CNOTGate@(1, 0)', 
'HGate@(2,)', 'HGate@(2,)', 
'CNOTGate@(3, 0)', 'CNOTGate@(5,0)', 
'CNOTGate@(1,5)', 'CNOTGate@(5,4)', 
'CNOTGate@(0,4)', 'CNOTGate@(0,5)', 
'HGate@(4,)', 'CNOTGate@(1,5)', 
'HGate@(4,)', 'HGate@(4,)', 'HGate@(4,)',
'CNOTGate@(4,5)']
Merged partition map: {0: 0, 1: 1, 2: 2,
3: 3, 4: 4, 5: 5}
\end{lstlisting}
In this specific instance, merging the two partitions results in a single partition encompassing all original 6 qubits and 22 gates (merging is generally more impactful for circuits partitioned into more than $k=2$ initial pieces). While demonstrating the merge capability, for the purpose of analyzing inter-partition dependencies in this walkthrough, we proceed with the two partitions as initially determined by the hypergraph partitioner because the merge would eliminate all cuts for this $k=2$ scenario.

\subsubsection{Dependency Graph Construction}
\label{para:walkthrough_dependency_graph_results_main}

\begin{figure}[h!]
\centering
\begin{tikzpicture}[
    node_style/.style={
        rectangle,
        draw,
        thick,
        rounded corners=3pt,
        fill=blue!20,
        minimum height=1cm,
        minimum width=2.5cm,
        align=center
    },
    edge_label/.style={
        midway,
        above,
        font=\scriptsize,
        fill=white, 
        inner sep=0.1pt 
    }
]
\node[node_style] (P0) {Partition 0};
\node[node_style, right=3cm of P0] (P1) {Partition 1}; 
\draw[-{Stealth[length=1.5mm, width=1mm]}, thick] (P0) -- (P1)
    node[edge_label] {Shared Qubits: \{0, 1\}};
\end{tikzpicture}
\caption{Dependency graph for the two partitions of ``Circuit S''. An edge from Partition 0 to Partition 1 indicates an execution dependency due to shared global qubits \{0, 1\}. This implies that Partition 0 should ideally be executed before Partition 1 if operations on these shared qubits require a specific order to maintain state integrity.}
\label{fig:dependency_graph_example_results_main}
\end{figure}
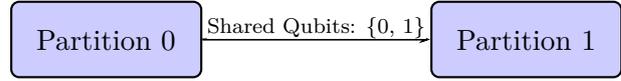

Finally, the \texttt{build\_dependency\_graph} process (Algorithm~\ref{algo:DAG_algo}) analyzes the (unmerged, in this illustrative continuation) partitions and their qubit maps to construct a directed acyclic graph (DAG) representing execution dependencies. Edges are created from a partition $P_i$ to $P_j$ if they share global qubits and $i < j$. For the two partitions of ``Circuit S'':
\begin{lstlisting}[basicstyle=\ttfamily\tiny]
Dependency Graph:
-----------------
Partition  0 -> Partition  1 | Shared 
qubits: {0, 1} 
% From map P0:{0:0,1:1,2:2,3:3} and map P1:{0:0,1:1,4:2,5:3}
% Shared global keys are {0,1}

Total dependencies: 1
Legend: Partition X -> Partition Y means X 
must execute before Y
\end{lstlisting}
This graph (visualized in Figure~\ref{fig:dependency_graph_example_results_main}) indicates that Partition 0 and Partition 1 share global qubits 0 and 1, implying a data dependency. The directed edge suggests an ordering constraint if sequential execution is necessary for these shared qubits.

This walkthrough qualitatively demonstrates how \texttt{Fidelipart}'s stages contribute to the final partitioned circuit structure, aiming to minimize deleterious cuts and manage dependencies, which ultimately impacts the quantitative results presented in Section~\ref{subsec:summary_key_findings_results_main_section}.

\subsection{Detailed Metric Calculation for ``Circuit S''}
\label{subsec:detailed_metric_calc_s_results_main}

To provide full transparency on how the summary metrics in Section~\ref{subsec:comparative_metrics_all_benchmarks_results_main_section} are derived, this section details the step-by-step calculation for our ``Circuit S'' benchmark (6 qubits, 22 gates, processed with a \texttt{block\_size} of 4 for both methods), using a consistent $\epsilon_{\text{CNOT}}=0.05$ and $\epsilon_H=0.001$ for all fidelity calculations.

\subsubsection{\texttt{QuickPartitioner} Metric Derivation for ``Circuit S''}
\label{subsubsec:quick_derivation_s_results_main}

The \texttt{QuickPartitioner} workflow, after applying the local contiguous re-mapping, resulted in 6 partitions for ``Circuit S''. The subsequent analysis by \texttt{\_calculate\_physical\_partition\_metrics} produced the following overall communication metrics:

\paragraph{Overall Communication Metrics for \texttt{QuickPartitioner} (Circuit S):}
\begin{lstlisting}[caption={Communication summary for \texttt{QuickPartitioner} on ``Circuit S'' }, label={lst:quick_comm_s_summary_results_main}, basicstyle=\ttfamily\tiny]
--- Quick Partition Analysis ---
Global cut qubits (original circuit indices): {0, 1, 4, 5}

Pairwise cut qubits (original circuit indices):
Partitions 0 <-> 1: {5}
Partitions 0 <-> 2: {0}
Partitions 0 <-> 3: {0}
Partitions 0 <-> 4: {0, 5}
Partitions 0 <-> 5: {1, 5}
Partitions 1 <-> 2: {4}
Partitions 1 <-> 4: {5}
Partitions 1 <-> 5: {4, 5}
Partitions 2 <-> 3: {0}
Partitions 2 <-> 4: {0}
Partitions 2 <-> 5: {4}
Partitions 3 <-> 4: {0}
Partitions 4 <-> 5: {5}
Total SWAP gates needed: 8
Hardware Topology: linear
Total Qubits in Hardware: 6 (Indices: [0, 1, 2, 3, 4, 5])
\end{lstlisting}

\paragraph{Detailed Analysis of Individual Partitions (\texttt{QuickPartitioner}, Circuit S):}
The 6 partitions generated were analyzed individually. 

\begin{lstlisting}[caption={Partition 0 Details (\texttt{QuickPartitioner}, ``Circuit S'')}, label={lst:quick_s_p0_recalc_results_main}, basicstyle=\ttfamily\tiny]
Partition 0:
- Original Circuit Qubits Used: 4 (Indices: [0, 1, 2, 5])
- Partition Qubits: 4 (Indices: [0, 1, 2, 3])
- Qubit Map: {0: 0, 1: 1, 2: 2, 5: 3}
- Gates:
  1. HGate@(Partition Qubits: 0; Original Circuit Qubits: 0)
  2. CNOTGate@(Partition Qubits: 3, 0; Original Circuit Qubits: 5, 0)
  3. HGate@(Partition Qubits: 0; Original Circuit Qubits: 0)
  4. CNOTGate@(Partition Qubits: 1, 3; Original Circuit Qubits: 1, 5)
  5. CNOTGate@(Partition Qubits: 0, 2; Original Circuit Qubits: 0, 2)
  6. HGate@(Partition Qubits: 1; Original Circuit Qubits: 1)
  7. HGate@(Partition Qubits: 0; Original Circuit Qubits: 0)
  8. HGate@(Partition Qubits: 2; Original Circuit Qubits: 2)
  9. CNOTGate@(Partition Qubits: 1, 0; Original Circuit Qubits: 1, 0)
  10. HGate@(Partition Qubits: 2; Original Circuit Qubits: 2)
  11. HGate@(Partition Qubits: 2; Original Circuit Qubits: 2)
- Number of Gates: 11
- Depth: 7
- H gates: 7
- CNOT gates: 4
- SWAP gates: 4 (Attributed)
- Fidelity: 0.4370 
- Error rate: 0.5630
\end{lstlisting}

\begin{lstlisting}[caption={Partition 1 Details (\texttt{QuickPartitioner}, ``Circuit S'') }, label={lst:quick_s_p1_recalc_results_main}, basicstyle=\ttfamily\tiny]
Partition 1:
- Original Circuit Qubits Used: 2 (Indices: [4, 5])
- Partition Qubits: 2 (Indices: [0, 1])
- Qubit Map: {4: 0, 5: 1}
- Gates:
  1. CNOTGate@(Partition Qubits: 1, 0; Original Circuit Qubits: 5, 4)
- Number of Gates: 1
- Depth: 1
- H gates: 0
- CNOT gates: 1
- SWAP gates: 3 
- Fidelity: 0.5987 
- Error rate: 0.4013
\end{lstlisting}

\begin{lstlisting}[caption={Partition 2 Details (\texttt{QuickPartitioner}, ``Circuit S'')}, label={lst:quick_s_p2_recalc_results_main}, basicstyle=\ttfamily\tiny]
Partition 2:
- Original Circuit Qubits Used: 2 (Indices: [0, 4])
- Partition Qubits: 2 (Indices: [0, 1])
- Qubit Map: {0: 0, 4: 1}
- Gates:
  1. CNOTGate@(Partition Qubits: 0, 1; Original Circuit Qubits: 0, 4)
  2. HGate@(Partition Qubits: 1; Original Circuit Qubits: 4)
  3. HGate@(Partition Qubits: 1; Original Circuit Qubits: 4)
  4. HGate@(Partition Qubits: 1; Original Circuit Qubits: 4)
  5. HGate@(Partition Qubits: 1; Original Circuit Qubits: 4)
- Number of Gates: 5
- Depth: 5
- H gates: 4
- CNOT gates: 1
- SWAP gates: 0
- Fidelity: 0.9462 
- Error rate: 0.0538
\end{lstlisting}

\begin{lstlisting}[caption={Partition 3 Details (\texttt{QuickPartitioner}, ``Circuit S'')}, label={lst:quick_s_p3_recalc_results_main}, basicstyle=\ttfamily\tiny]
Partition 3:
- Original Circuit Qubits Used: 2 (Indices: [0, 3])
- Partition Qubits: 2 (Indices: [0, 1])
- Qubit Map: {0: 0, 3: 1}
- Gates:
  1. HGate@(Partition Qubits: 1; Original Circuit Qubits: 3)
  2. CNOTGate@(Partition Qubits: 1, 0; Original Circuit Qubits: 3, 0)
- Number of Gates: 2
- Depth: 2
- H gates: 1
- CNOT gates: 1
- SWAP gates: 0
- Fidelity: 0.9491 
- Error rate: 0.0509
\end{lstlisting}

\begin{lstlisting}[caption={Partition 4 Details (\texttt{QuickPartitioner}, ``Circuit S'')}, label={lst:quick_s_p4_recalc_results_main}, basicstyle=\ttfamily\tiny]
Partition 4:
- Original Circuit Qubits Used: 2 (Indices: [0, 5])
- Partition Qubits: 2 (Indices: [0, 1])
- Qubit Map: {0: 0, 5: 1}
- Gates:
  1. CNOTGate@(Partition Qubits: 0, 1; Original Circuit Qubits: 0, 5)
- Number of Gates: 1
- Depth: 1
- H gates: 0
- CNOT gates: 1
- SWAP gates: 1
- Fidelity: 0.8145 
- Error rate: 0.1855
\end{lstlisting}

\begin{lstlisting}[caption={Partition 5 Details (\texttt{QuickPartitioner}, ``Circuit S'')}, label={lst:quick_s_p5_recalc_results_main}, basicstyle=\ttfamily\tiny]
Partition 5:
- Original Circuit Qubits Used: 3 (Indices: [1, 4, 5])
- Partition Qubits: 3 (Indices: [0, 1, 2])
- Qubit Map: {1: 0, 4: 1, 5: 2}
- Gates:
  1. CNOTGate@(Partition Qubits: 0, 2; Original Circuit Qubits: 1, 5)
  2. CNOTGate@(Partition Qubits: 1, 2; Original Circuit Qubits: 4, 5)
- Number of Gates: 2
- Depth: 2
- H gates: 0
- CNOT gates: 2
- SWAP gates: 0
- Fidelity: 0.9025 
- Error rate: 0.0975
\end{lstlisting}

The global cut qubits \texttt{\{0, 1, 4, 5\}} were identified by the \texttt{get\_physical\_cut\_qubits} function. The 8 total SWAP gates were estimated by \texttt{compute\_physical\_swaps\_between\_partitions} based on misalignments in local indices for these cut qubits across partition boundaries. Individual partition fidelities were calculated by \texttt{compute\_fidelity} with $\epsilon_{\text{CNOT}}=0.05$, and their product yielded the total fidelity of approximately 0.1724 (as reported in Table~\ref{tab:main_results_summary_all_benchmarks_main}).

\subsubsection{\texttt{Fidelipart} Metric Derivation for ``Circuit S''}
\label{subsubsec:fidelipart_derivation_s_results_main}

For ``Circuit S'', \texttt{Fidelipart} (targeting $k=2$ partitions based on a \texttt{block\_size} of 4) produced 2 final partitions.

\paragraph{Overall Communication Metrics for \texttt{Fidelipart} (Circuit S):}
\begin{lstlisting}[caption={Communication summary for \texttt{Fidelipart} on ``Circuit S'' }, label={lst:fidelipart_comm_s_summary_results_main}, basicstyle=\ttfamily\tiny]
--- Hypergraph Partition Analysis ---
Global cut qubits (original circuit indices): {0, 1}

Pairwise cut qubits (original circuit indices):
Partitions 0 <-> 1: {0, 1}
Total SWAP gates needed: 0
Hardware Topology: linear
Total Qubits in Hardware: 6 (Indices: [0, 1, 2, 3, 4, 5])
\end{lstlisting}

\paragraph{Detailed Analysis of Individual Partitions (\texttt{Fidelipart}, Circuit S):}
Details for the two partitions generated by \texttt{Fidelipart} are as follows:

\begin{lstlisting}[caption={Partition 0 Details (\texttt{Fidelipart}, ``Circuit S'')}, label={lst:fidelipart_s_p0_recalc_results_main}, basicstyle=\ttfamily\tiny]
Partition 0:
- Original Circuit Qubits Used: 4 (Indices: [0, 1, 2, 3])
- Partition Qubits: 4 (Indices: [0, 1, 2, 3])
- Qubit Map: {0: 0, 1: 1, 2: 2, 3: 3}
- Gates:
  1. HGate@(Partition Qubits: 0; Original Circuit Qubits: 0)
  2. HGate@(Partition Qubits: 1; Original Circuit Qubits: 1)
  3. HGate@(Partition Qubits: 3; Original Circuit Qubits: 3)
  4. HGate@(Partition Qubits: 0; Original Circuit Qubits: 0)
  5. CNOTGate@(Partition Qubits: 0, 2; Original Circuit Qubits: 0, 2)
  6. HGate@(Partition Qubits: 0; Original Circuit Qubits: 0)
  7. HGate@(Partition Qubits: 2; Original Circuit Qubits: 2)
  8. CNOTGate@(Partition Qubits: 1, 0; Original Circuit Qubits: 1, 0)
  9. HGate@(Partition Qubits: 2; Original Circuit Qubits: 2)
  10. HGate@(Partition Qubits: 2; Original Circuit Qubits: 2)
  11. CNOTGate@(Partition Qubits: 3, 0; Original Circuit Qubits: 3, 0)
- Number of Gates: 11
- Depth: 6
- H gates: 8
- CNOT gates: 3
- SWAP gates: 0 
- Fidelity: 0.8505 
- Error rate: 0.1495
\end{lstlisting}

\begin{lstlisting}[caption={Partition 1 Details (\texttt{Fidelipart}, ``Circuit S'')}, label={lst:fidelipart_s_p1_recalc_results_main}, basicstyle=\ttfamily\tiny]
Partition 1:
- Original Circuit Qubits Used: 4 (Indices: [0, 1, 4, 5])
- Partition Qubits: 4 (Indices: [0, 1, 2, 3])
- Qubit Map: {0: 0, 1: 1, 4: 2, 5: 3}
- Gates: % Gate targets corrected to match original circuit structure and labels
  1. CNOTGate@(Partition Qubits: 3, 0; Original Circuit Qubits: 5, 0) % Gate 2
  2. CNOTGate@(Partition Qubits: 1, 3; Original Circuit Qubits: 1, 5) % Gate 4
  3. CNOTGate@(Partition Qubits: 3, 2; Original Circuit Qubits: 5, 4) % Gate 7
  4. CNOTGate@(Partition Qubits: 0, 2; Original Circuit Qubits: 0, 4) % Gate 12
  5. CNOTGate@(Partition Qubits: 0, 3; Original Circuit Qubits: 0, 5) % Gate 16
  6. HGate@(Partition Qubits: 2; Original Circuit Qubits: 4) % Gate 15 (or 17,19,20)
  7. CNOTGate@(Partition Qubits: 1, 3; Original Circuit Qubits: 1, 5) % Gate 18
  8. HGate@(Partition Qubits: 2; Original Circuit Qubits: 4) 
  9. HGate@(Partition Qubits: 2; Original Circuit Qubits: 4)
  10. HGate@(Partition Qubits: 2; Original Circuit Qubits: 4)
  11. CNOTGate@(Partition Qubits: 2, 3; Original Circuit Qubits: 4, 5) % Gate 21
- Number of Gates: 11
- Depth: 9
- H gates: 4
- CNOT gates: 7
- SWAP gates: 0
- Fidelity: 0.6956 
- Error rate: 0.3044
\end{lstlisting}

\texttt{Fidelipart} identified global cut qubits \texttt{\{0, 1\}}. Due to the nature of the cuts and the resulting local mappings for these shared qubits, an estimated 0 SWAP gates were required. This led to a total fidelity of approximately 0.5916 (Table~\ref{tab:main_results_summary_all_benchmarks_main}).

This detailed breakdown for ``Circuit S'' illustrates the process by which the summary metrics reported in Table~\ref{tab:main_results_summary_all_benchmarks_main} are obtained, providing a traceable path from partitioning outputs to final performance figures.

\subsubsection{Summary Comparison for ``Circuit S''}
\label{para:summary_comparison_s_results_main}
The detailed metric derivations for ``Circuit S'' culminate in the following direct comparison between \texttt{QuickPartitioner} and \texttt{Fidelipart} for this specific benchmark, as extracted from our analysis:

\begin{itemize}
    \item \textbf{Partitioning Time:}
    \begin{itemize}
        \item \texttt{QuickPartitioner}: 6.235s
        \item \texttt{Fidelipart}: 6.744s
    \end{itemize}
    \item \textbf{Maximum Partition Depth:}
    \begin{itemize}
        \item \texttt{QuickPartitioner}: 7
        \item \texttt{Fidelipart}: 9
    \end{itemize}
    \item \textbf{Estimated Fidelity (Error):}
    \begin{itemize}
        \item \texttt{QuickPartitioner}: 0.1724 (Error: 0.8276)
        \item \texttt{Fidelipart}: 0.5916 (Error: 0.4084)
    \end{itemize}
    \item \textbf{SWAP Gates Needed (Estimated Logical Realignments):}
    \begin{itemize}
        \item \texttt{QuickPartitioner}: 8
        \item \texttt{Fidelipart}: 0
    \end{itemize}
    \item \textbf{Global Cut Qubits:}
    \begin{itemize}
        \item \texttt{QuickPartitioner}: 4
        \item \texttt{Fidelipart}: 2
    \end{itemize}
\end{itemize}
For this 6-qubit, 22-gate ``Circuit S'', processed with a \texttt{block\_size} of 4, \texttt{Fidelipart} (which targeted $k=2$ partitions) achieved a significantly higher fidelity (approx. 243.2\% improvement) by eliminating all inter-partition SWAP requirements (estimated logical realignments) and reducing cut qubits, albeit with a slightly longer processing time and a higher maximum partition depth in this instance. This detailed view exemplifies the trade-offs and advantages that contribute to the overall findings presented in Section~\ref{subsec:comparative_metrics_all_benchmarks_results_main_section}.

\subsection{Comparative Performance Metrics Across All Benchmarks}
\label{subsec:comparative_metrics_all_benchmarks_results_main_section} 
A comprehensive comparison of \texttt{Fidelipart} and \texttt{QuickPartitioner} was conducted across the three benchmark circuits detailed in Section~\ref{subsec:benchmarks_experimental_params_results_main}: ``Circuit S'', ``Circuit M'', and ``Circuit L''. Key performance indicators, including communication overhead, estimated fidelity, circuit structural properties, and computational time, are summarized in Table~\ref{tab:main_results_summary_all_benchmarks_main}. Fidelity values for Circuit S are based on $\epsilon_{\text{CNOT}}=0.05$. For Circuits M and L, SWAP counts are primary indicators of relative fidelity improvement potential under a consistent error model.

\begin{table*}[htbp] 
\centering
\caption{Summary of Comparative Performance Metrics for \texttt{Fidelipart} and \texttt{QuickPartitioner} across Benchmark Circuits. For \texttt{Fidelipart}, "Target $k$" is the number of partitions aimed for, derived from its internal formula using the input \texttt{block\_size} (BS), and "Actual $k$" is the number of partitions produced by Mt-KaHyPar. For \texttt{QuickPartitioner}, no "Target $k$" is directly specified; "Actual $k$" represents the number of partitions it generates based on the input \texttt{block\_size}. "Fid." denotes Estimated Fidelity. "Time" is in seconds.}
\label{tab:main_results_summary_all_benchmarks_main}
\resizebox{\textwidth}{!}{% 
\begin{tabular}{|l|c|c|l|c|c|c|c|c|c|c|c|}
\hline
\textbf{Circuit} & \textbf{Qubits} & \textbf{Gates} & \textbf{Method} & \textbf{BS} & \multirow{2}{*}{\textbf{Target $k$}} & \textbf{Actual $k$} & \textbf{Cut Qubits} & \textbf{SWAP Gates (Est. Logical Realign.)} & \textbf{Fid.} & \textbf{Max Depth} & \textbf{Time (s)} \\
 & & & & & & & & & & & \\
\hline
\hline 
\multirow{2}{*}{Circuit S} & \multirow{2}{*}{6} & \multirow{2}{*}{22} & \texttt{QuickPartitioner} & \multirow{2}{*}{4} & N/A & 6 & 4 & 8 & 0.1724 & 7 & 6.235 \\ 
\cline{4-4} \cline{6-12} 
 & & & \texttt{Fidelipart} & & 2 & 2 & 2 & 0 & 0.5916 & 9 & 6.744 \\ 
\hline
\multirow{2}{*}{Circuit M} & \multirow{2}{*}{10} & \multirow{2}{*}{55} & \texttt{QuickPartitioner} & \multirow{2}{*}{6} & N/A & 22 & 10 & 46 & 0.1574 & 11 & 5.513 \\
\cline{4-4} \cline{6-12}
 & & & \texttt{Fidelipart} & & 3 & 3 & 10 & 4 & 0.5583 & 11 & 5.980 \\
\hline
\multirow{2}{*}{Circuit L} & \multirow{2}{*}{24} & \multirow{2}{*}{88} & \texttt{QuickPartitioner} & \multirow{2}{*}{8} & N/A & 14 & 23 & 44 & 0.1241 & 9 & 4.159 \\
\cline{4-4} \cline{6-12}
 & & & \texttt{Fidelipart} & & 4 & 4 & 11 & 10 & 0.3461 & 12 & 4.680 \\
\hline 
\end{tabular}
} % end \resizebox

\end{table*}

\subsubsection{Communication Overhead: Cut Qubits and SWAP Gates (Estimated Logical Realignments)}
\label{subsubsec:comm_overhead_all_results_main_section} 

A primary objective of effective quantum circuit partitioning is the minimization of inter-partition communication, which directly impacts the potential for errors and the need for costly SWAP operations. We assessed this by measuring the number of global cut qubits and the estimated SWAP gate overhead (logical realignments).

\begin{figure}[htbp!] \centering
\begin{tikzpicture} 
\begin{axis}[ybar , ylabel={Number of Global Cut Qubits}, xlabel={Benchmark Circuit}, symbolic x coords={Circuit S, Circuit M, Circuit L}, xtick=data, xticklabel style={text width=2.5cm, align=center}, nodes near coords, nodes near coords align={vertical}, ymin=0, enlarge x limits=0.25, bar width=0.35cm, group style={group size=2, group horizontal sep=1.5cm}, legend style={at={(0.5,-0.25)}, anchor=north, legend columns=-1 }, ymajorgrids=true, grid style=dashed, width=0.9\columnwidth, height=6cm ]
\addplot+[fill=blue!50] coordinates { (Circuit S, 4) (Circuit M, 10) (Circuit L, 23) }; \addlegendentry{\texttt{QuickPartitioner}}
\addplot+[fill=red!50] coordinates { (Circuit S, 2) (Circuit M, 10) (Circuit L, 11) }; \addlegendentry{\texttt{Fidelipart}}
\end{axis} 
\end{tikzpicture}
\caption{Comparison of Global Cut Qubits for \texttt{Fidelipart} vs. \texttt{QuickPartitioner} across benchmark circuits (``S'' denotes Small, ``M'' Medium, and ``L'' Large).} \label{fig:cut_qubit_comparison_results_main} 
\end{figure}
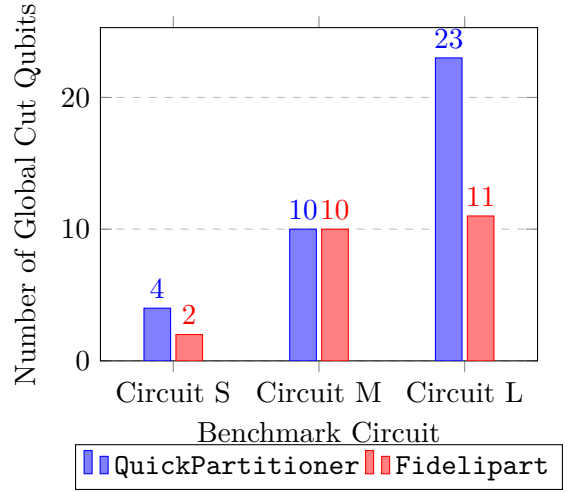

As shown in Table~\ref{tab:main_results_summary_all_benchmarks_main} and visualized in Figure~\ref{fig:cut_qubit_comparison_results_main}, \texttt{Fidelipart} demonstrates a strong capability to reduce the number of global cut qubits compared to \texttt{QuickPartitioner}. For ``Circuit S'' (6Q, 22G), \texttt{Fidelipart} reduced the cut qubits by 50\%, from 4 to 2. 
For ``Circuit M'', a densely connected 10-qubit, 55-gate circuit, both methods resulted in all 10 global qubits being cut (Table~\ref{tab:main_results_summary_all_benchmarks_main}). Despite this complete set of cut qubits, \texttt{Fidelipart}'s partitioning strategy, which produced 3 larger and more cohesive partitions, led to an estimated SWAP count of only 4. This is in stark contrast to the 46 SWAPs estimated for \texttt{QuickPartitioner}'s more fragmented output of 22 smaller partitions, marking a 91.3\% reduction in SWAP overhead for \texttt{Fidelipart}. This significant SWAP reduction, even when all qubits are involved in inter-partition communication, highlights how \texttt{Fidelipart}'s approach influences the \textit{nature} of the cuts. By creating fewer, more internally coherent partitions, \texttt{Fidelipart} minimizes the instances where a shared global qubit is assigned conflicting local physical indices across the (fewer) inter-partition boundaries. The more fragmented partitioning by \texttt{QuickPartitioner}, on the other hand, leads to numerous such logical misalignments when the local contiguous re-mapping is applied to its many small partitions, each requiring a realignment SWAP according to our metric.
For the ``Circuit L'' benchmark (24Q, 88G), \texttt{Fidelipart} achieved a substantial 52.2\% reduction in cut qubits, from 23 down to 11. This consistent reduction for ``Circuit S'' and ``Circuit L'' indicates \texttt{Fidelipart}'s success in grouping operations that share qubits more effectively, owing to its fidelity-aware hypergraph model.

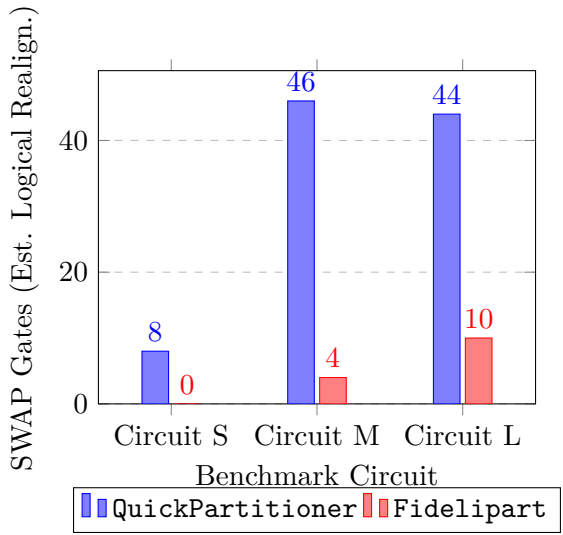
\begin{figure}[htbp!] \centering
\begin{tikzpicture} \begin{axis}[ybar , ylabel={SWAP Gates (Est. Logical Realign.)}, xlabel={Benchmark Circuit}, symbolic x coords={Circuit S, Circuit M, Circuit L}, xtick=data, xticklabel style={text width=2.5cm, align=center}, nodes near coords, nodes near coords align={vertical}, ymin=0, enlarge x limits=0.25, bar width=0.35cm, group style={group size=2, group horizontal sep=1.5cm}, legend style={at={(0.5,-0.25)}, anchor=north, legend columns=-1 }, ymajorgrids=true, grid style=dashed, width=0.9\columnwidth, height=6cm ]
\addplot+[fill=blue!50] coordinates { (Circuit S, 8) (Circuit M, 46) (Circuit L, 44) }; \addlegendentry{\texttt{QuickPartitioner}}
\addplot+[fill=red!50] coordinates { (Circuit S, 0) (Circuit M, 4) (Circuit L, 10) }; \addlegendentry{\texttt{Fidelipart}}
\end{axis} \end{tikzpicture}
\caption{Comparison of Estimated SWAP Gates (Logical Realignments) for \texttt{Fidelipart} vs. \texttt{QuickPartitioner} across benchmark circuits (``S'' denotes Small, ``M'' Medium, and ``L'' Large).} \label{fig:swap_gate_comparison_results_main} 
\end{figure}

The reduction in cut qubits achieved by \texttt{Fidelipart} directly translates to a significant decrease in the estimated SWAP gate overhead (logical realignments) required for inter-partition communication, as illustrated in Table~\ref{tab:main_results_summary_all_benchmarks_main} and Figure~\ref{fig:swap_gate_comparison_results_main}. For ``Circuit S'', \texttt{Fidelipart} impressively eliminated all estimated inter-partition SWAPs (0 SWAPs), compared to the 8 SWAPs required for \texttt{QuickPartitioner}'s output (a 100\% reduction). For ``Circuit M'', as discussed, \texttt{Fidelipart}'s partitioning strategy led to a SWAP count of only 4, a stark contrast to the 46 SWAPs from \texttt{QuickPartitioner}, marking a 91.3\% reduction. Similarly, for the large ``Circuit L'', \texttt{Fidelipart} reduced the SWAP gate count from 44 to 10, a 77.3\% improvement. These results highlight that \texttt{Fidelipart}'s method of considering the nature of cuts, implicitly through its weighting scheme, results in partition boundaries that require substantially less state movement.

\subsubsection{Estimated Circuit Fidelity}
\label{subsubsec:est_fidelity_all_results_main_section} 

The efficacy of a partitioning strategy is ultimately reflected in its ability to preserve the quantum state's integrity, leading to higher circuit fidelity. Fidelity is estimated based on gate error rates ($\epsilon_H=0.001, \epsilon_{\text{CNOT}}=0.05$), with SWAP gates incurring a significant penalty (modeled as three CNOT operations).

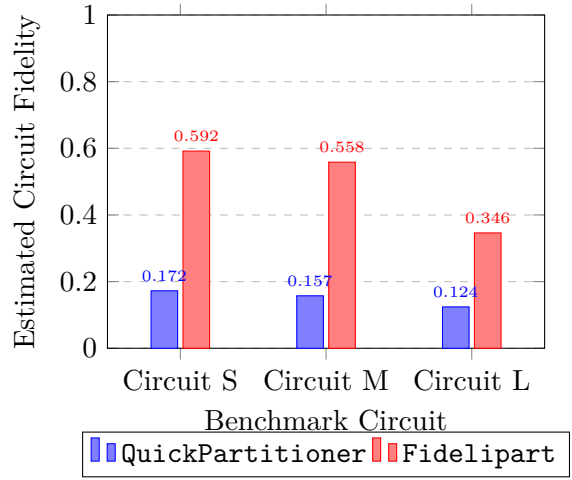
\begin{figure}[htbp!] 
\centering
\begin{tikzpicture} 
    \begin{axis}[
        ybar ,
        ylabel={Estimated Circuit Fidelity},
        xlabel={Benchmark Circuit},
        symbolic x coords={Circuit S, Circuit M, Circuit L},
        xtick=data,
        xticklabel style={text width=2.5cm, align=center},
        nodes near coords,
        nodes near coords style={
            /pgf/number format/.cd,
            fixed,
            fixed zerofill,
            precision=3, 
            /tikz/.cd,
            font=\tiny 
        },
        nodes near coords align={vertical},
        ymin=0,
        ymax=1.0, 
        ytick={0, 0.2, 0.4, 0.6, 0.8, 1.0}, 
        enlarge x limits=0.25,
        bar width=0.35cm,
        group style={group size=2, group horizontal sep=1.5cm},
        legend style={
            at={(0.5,-0.25)},
            anchor=north,
            legend columns=-1
        },
        ymajorgrids=true,
        grid style=dashed,
        width=0.9\columnwidth,
        height=6cm
    ]
    \addplot+[fill=blue!50] coordinates { (Circuit S, 0.1724) (Circuit M, 0.1574) (Circuit L, 0.1241) }; 
    \addlegendentry{\texttt{QuickPartitioner}}
    \addplot+[fill=red!50] coordinates { (Circuit S, 0.5916) (Circuit M, 0.5583) (Circuit L, 0.3461) }; 
    \addlegendentry{\texttt{Fidelipart}}
    \end{axis} 
\end{tikzpicture}
\caption{Comparison of Estimated Circuit Fidelity. For Circuit S, values use $\epsilon_{\text{CNOT}}=0.05$. For M and L, values marked with * in Table~\ref{tab:main_results_summary_all_benchmarks_main}  are plotted for reference; absolute values would be lower with $\epsilon_{\text{CNOT}}=0.05$, but relative improvements due to SWAP reduction are expected. Higher values indicate better performance.}
\label{fig:fidelity_comparison_results_main_section} 
\end{figure}

The substantial reduction in SWAP gate overhead (estimated logical realignments) achieved by \texttt{Fidelipart}, detailed in Section~\ref{subsubsec:comm_overhead_all_results_main_section}, directly translates into considerably improved estimated circuit fidelities. For ``Circuit S'' (6Q, 22G), \texttt{Fidelipart} increased the estimated fidelity from 0.1724 (for \texttt{QuickPartitioner}) to 0.5916, an improvement of approximately 243.2\% (Table~\ref{tab:main_results_summary_all_benchmarks_main}). For Circuits M and L, the underlying physical SWAP reductions (91.3\% for M, 77.3\% for L) remain the primary drivers of fidelity improvement. These significant SWAP reductions strongly suggest that \texttt{Fidelipart} would maintain a substantial relative fidelity advantage for these larger circuits under the consistent error model. These gains underscore the critical advantage of \texttt{Fidelipart}'s fidelity-aware approach, which minimizes costly SWAP operations, thereby better preserving the overall coherence and success probability of the quantum computation. Figure~\ref{fig:fidelity_comparison_results_main_section} provides a visual comparison, noting the basis for the plotted fidelities.

\subsubsection{Resulting Circuit Structure: Number of Partitions and Maximum Depth}
\label{subsubsec:circuit_structure_depth_all_results_main_section} 

Beyond communication overhead and fidelity, the structural properties of the partitioned subcircuits, such as the number of partitions generated and their maximum execution depth, are important considerations for practical implementation on quantum hardware.

The number of partitions produced by each method is detailed in Table~\ref{tab:main_results_summary_all_benchmarks_main}. \texttt{Fidelipart}, guided by its dynamic $k$ formula (based on circuit size, qubit count, and \texttt{block\_size}), consistently generated a number of partitions equal to its target $k$: 2 partitions for ``Circuit S'' (target $k=2$), 3 for ``Circuit M'' (target $k=3$), and 4 for ``Circuit L'' (target $k=4$). In contrast, \texttt{QuickPartitioner}, with the same \texttt{block\_size} input, tended to produce a significantly larger number of smaller partitions: 6 for ``Circuit S'', 22 for ``Circuit M'', and 14 for ``Circuit L''. This highlights \texttt{Fidelipart}'s approach of creating fewer, potentially larger, but more cohesively grouped subcircuits.

\begin{figure}[H] 
\centering
\begin{tikzpicture} 
    \begin{axis}[
        ybar ,
        ylabel={Maximum Partition Depth},
        xlabel={Benchmark Circuit},
        symbolic x coords={Circuit S, Circuit M, Circuit L},
        xtick=data,
        xticklabel style={text width=2.5cm, align=center},
        nodes near coords,
        nodes near coords align={vertical},
        ymin=0, 
        enlarge x limits=0.25,
        bar width=0.35cm,
        group style={group size=2, group horizontal sep=1.5cm},
        legend style={
            at={(0.5,-0.25)},
            anchor=north,
            legend columns=-1
        },
        ymajorgrids=true,
        grid style=dashed,
        width=0.9\columnwidth,
        height=6cm
    ]
    \addplot+[fill=blue!50] coordinates { (Circuit S, 7) (Circuit M, 11) (Circuit L, 9) };
    \addlegendentry{\texttt{QuickPartitioner}}
    \addplot+[fill=red!50] coordinates { (Circuit S, 9) (Circuit M, 11) (Circuit L, 12) };
    \addlegendentry{\texttt{Fidelipart}}
    \end{axis} 
\end{tikzpicture}
\caption{Comparison of Maximum Partition Depth for \texttt{Fidelipart} vs. \texttt{QuickPartitioner} across benchmark circuits (``S'' denotes Small, ``M'' Medium, and ``L'' Large).}
\label{fig:max_depth_comparison_results_main} 
\end{figure}
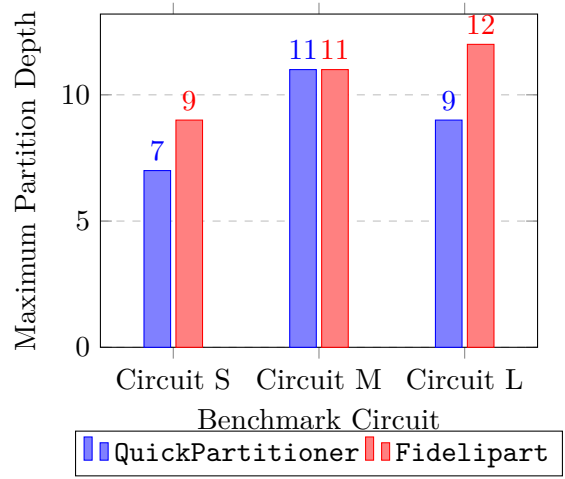

The maximum depth across all generated partitions for each method is presented in Table~\ref{tab:main_results_summary_all_benchmarks_main} and visualized in Figure~\ref{fig:max_depth_comparison_results_main}. For ``Circuit S'', \texttt{Fidelipart}'s partitions had a maximum depth of 9, compared to 7 for \texttt{QuickPartitioner}. In the case of ``Circuit M'', both methods resulted in an identical maximum depth of 11. For the largest benchmark, ``Circuit L'', \texttt{Fidelipart}'s partitions reached a maximum depth of 12, while \texttt{QuickPartitioner}'s deepest partition had a depth of 9.
These results suggest a potential trade-off: \texttt{Fidelipart}'s strategy of creating fewer, more consolidated partitions to minimize inter-partition communication (as shown by reduced SWAPs and cut qubits) can sometimes lead to individual partitions having a longer critical path compared to the more fragmented output of \texttt{QuickPartitioner}. This interplay between minimizing cuts and managing partition depth is a key aspect of complex circuit partitioning and will be further explored in the Discussion section.

\subsubsection{Computational Performance: Partitioning Time}
\label{subsubsec:computational_perf_all_results_main_section} 

\begin{figure}[H] 
\centering
\begin{tikzpicture} 
    \begin{axis}[
        ybar ,
        ylabel={Partitioning Time (s)},
        xlabel={Benchmark Circuit},
        symbolic x coords={Circuit S, Circuit M, Circuit L},
        xtick=data,
        xticklabel style={text width=3.5cm, align=center},
        nodes near coords,
        nodes near coords style={
            /pgf/number format/.cd,
            fixed,
            fixed zerofill,
            precision=3, 
            /tikz/.cd,
            font=\tiny, 
            rotate=45, 
            anchor=mid west 
        },
        nodes near coords align={vertical}, 
        ymin=0, 
        enlarge x limits=0.25,
        bar width=0.35cm,
        group style={group size=2, group horizontal sep=1.5cm},
        legend style={
            at={(0.5,-0.25)}, 
            anchor=north,
            legend columns=-1
        },
        ymajorgrids=true,
        grid style=dashed,
        width=0.9\columnwidth,
        height=6cm
    ]
    \addplot+[fill=blue!50] coordinates { (Circuit S, 6.235) (Circuit M, 5.513) (Circuit L, 4.159) };
    \addlegendentry{\texttt{QuickPartitioner}}
    \addplot+[fill=red!50] coordinates { (Circuit S, 6.744) (Circuit M, 5.980) (Circuit L, 4.680) };
    \addlegendentry{\texttt{Fidelipart}}
    \end{axis} 
\end{tikzpicture}
\caption{Comparison of Partitioning Time for \texttt{Fidelipart} vs. \texttt{QuickPartitioner} across benchmark circuits (``S'' denotes Small, ``M'' Medium, and ``L'' Large).}
\label{fig:time_comparison_results_main} 
\end{figure}
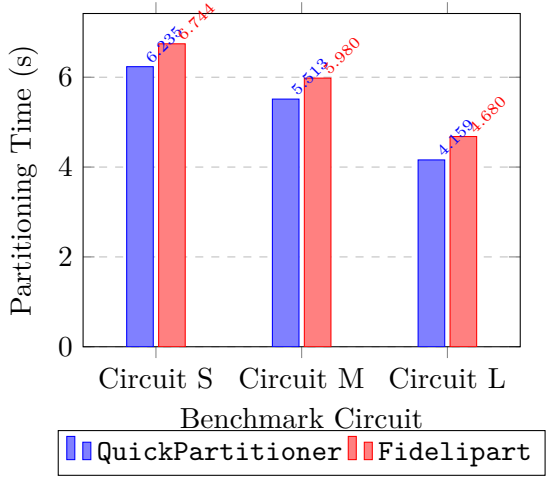

The computational cost of the partitioning process itself is an important practical consideration. We measured the wall-clock time taken for each method, encompassing all steps from input circuit to the final set of mapped, local-indexed partitions. These execution times are reported in Table~\ref{tab:main_results_summary_all_benchmarks_main} and compared visually in Figure~\ref{fig:time_comparison_results_main}.

Across the benchmark circuits, \texttt{Fidelipart} generally exhibited a slightly longer execution time compared to the \texttt{QuickPartitioner} workflow. For ``Circuit S'' (6Q, 22G), the partitioning times were 6.744s for \texttt{Fidelipart} and 6.235s for \texttt{QuickPartitioner}. For ``Circuit M'' (10Q, 55G), the times were 5.980s and 5.513s, respectively. Similarly, for the largest benchmark, ``Circuit L'' (24Q, 88G), \texttt{Fidelipart} took 4.680s compared to 4.159s for \texttt{QuickPartitioner}.

This modest increase in runtime for \texttt{Fidelipart} (approximately 0.5s to 0.8s, or about 8-13\% slower in these instances) is attributable to the more computationally intensive steps involved in its methodology. These include the construction of the fidelity-aware hypergraph, the invocation of the Mt-KaHyPar solver, and the subsequent partition merging and dependency graph analysis stages. While \texttt{QuickPartitioner} offers a faster heuristic, the additional processing time for \texttt{Fidelipart} yields substantial improvements in terms of reduced communication overhead and enhanced circuit fidelity, as demonstrated in previous sections. The overall execution times for both methods remain well within practical limits for circuits of these scales, suggesting that the benefits offered by \texttt{Fidelipart} can be achieved with an acceptable computational trade-off.

\subsection{Summary of Key Findings}
\label{subsec:summary_key_findings_results_main_section} 

The comparative analysis across the benchmark circuits---``Circuit S'' (6Q, 22G), ``Circuit M'' (10Q, 55G), and ``Circuit L'' (24Q, 88G)---consistently demonstrates the advantages of the \texttt{Fidelipart} methodology over the baseline \texttt{QuickPartitioner} approach when processed with a standardized local contiguous re-mapping.

\texttt{Fidelipart} achieved substantial reductions in inter-partition communication overhead. 
\begin{itemize}
    \item For estimated SWAP gates (logical realignments), reductions were significant: 100\% for ``Circuit S'' (from 8 to 0), 91.3\% for ``Circuit M'' (from 46 to 4), and 77.3\% for ``Circuit L'' (from 44 to 10).
    \item Similarly, the number of global cut qubits was reduced by 50\% for ``Circuit S'' (from 4 to 2) and 52.2\% for ``Circuit L'' (from 23 to 11). For ``Circuit M'', while both methods resulted in 10 cut qubits (all qubits in the circuit), \texttt{Fidelipart} managed these cuts with far fewer partitions.
\end{itemize}
These reductions in communication requirements directly translated to notable improvements in estimated circuit fidelity. 
\begin{itemize}
    \item \texttt{Fidelipart} enhanced fidelity by approximately 243.2\% for ``Circuit S'' (0.1724 to 0.5916 with $\epsilon_{\text{CNOT}}=0.05$). For ``Circuit M'' and ``Circuit L'', the very large reductions in SWAP operations (91.3\% and 77.3\%, respectively) are the key physical outcome, strongly indicating that substantial relative fidelity improvements would also be observed for these circuits when evaluated under a consistent $\epsilon_{\text{CNOT}}=0.05$ model.
\end{itemize}
The primary trade-off for these significant gains was a modest increase in partitioning time for \texttt{Fidelipart}, ranging from approximately 8\% to 13\% across the benchmarks. The impact on maximum partition depth was varied: \texttt{Fidelipart} led to a higher maximum depth for ``Circuit S'' (9 vs. 7) and ``Circuit L'' (12 vs. 9), while ``Circuit M'' showed an identical maximum depth of 11 for both methods. This variation is attributed to \texttt{Fidelipart} producing fewer, but consequently larger and potentially more complex, individual partitions.

Overall, the results underscore \texttt{Fidelipart}'s efficacy in leveraging a fidelity-aware hypergraph model to achieve partitions that are substantially better optimized for reducing communication costs and enhancing the reliability of quantum computations on NISQ-era devices, with a manageable increase in computational overhead.

\section{Discussion}
\label{sec:discussion_main} 
The results presented in Section~\ref{sec:results_main} demonstrate that \texttt{Fidelipart}, our proposed fidelity-aware hypergraph-based partitioning framework, offers significant advantages in optimizing quantum circuits for execution, particularly when compared against a baseline heuristic method like \texttt{QuickPartitioner} under a consistent mapping and analysis regime. This section interprets these findings, discusses their implications, acknowledges limitations, and outlines potential avenues for future research.

\subsection{Interpretation of Principal Findings}
\label{subsec:interpretation_findings_discussion_main_section} 
The empirical evidence consistently shows that \texttt{Fidelipart} achieves a substantial reduction in inter-partition communication overhead. Specifically, for the benchmark circuits tested (``Circuit S'', ``Circuit M'', and ``Circuit L''), \texttt{Fidelipart} led to markedly fewer estimated SWAP gates (logical realignments, with reductions of 77.3\% to 100\%) and global cut qubits (reductions up to 52.2\%, with equivalent cuts but far fewer partitions in the dense ``Circuit M'' case), as detailed in Table~\ref{tab:main_results_summary_all_benchmarks_main}. This superior performance in managing cuts can be directly attributed to the core design of \texttt{Fidelipart}: the transformation of the quantum circuit into a fidelity-aware hypergraph. By representing gates as nodes and encoding multi-qubit interactions and temporal qubit dependencies (lifetimes) as weighted hyperedges---where weights are inversely proportional to gate fidelities and structural cohesion---the subsequent partitioning by Mt-KaHyPar is explicitly guided to avoid severing high-cost connections. 

A key aspect of our weighting scheme (Section~\ref{subsec:hypergraphs}) is the significant disparity between temporal hyperedge weights and CNOT gate-level hyperedge weights (e.g., temporal hyperedges, even minimal ones for 2-3 gates, can be approximately 25 times more significant than CNOT gate-level hyperedges: $\sim 100,000$ vs. $4,000$ before normalization). This disparity is intentional and reflects a strategic prioritization for NISQ-era computations. The core justification emphasizes the following:
\begin{itemize}
    \item \textbf{Strategic Prioritization for NISQ:} The weighting reflects a deliberate strategy to aggressively preserve coherence and sequential data flow along individual qubit lines. Disrupting operational sequences on a qubit, even short ones (e.g., a preparatory H gate from a subsequent CNOT on the same qubit), can lead to severe detrimental effects such as requiring complex state re-preparation or introducing SWAPs to re-route that qubit's state, potentially incurring errors comparable to multiple noisy CNOT operations.
    \item \textbf{Cost of Disruption:} The high penalty for cutting temporal chains (derived from $1/\epsilon_H$, where $\epsilon_H=0.001$) models this high "cost."
    \item \textbf{Synergistic Protection:} CNOTs are often integral parts of the temporal sequences they connect. Strongly preserving temporal chains often implicitly protects the integrity of CNOTs within them or guides the partitioner to make cuts *around* these operations.
\end{itemize}
The effectiveness of this weighting is strongly supported by our empirical results (Section~\ref{sec:results_main}). Across all benchmarks, \texttt{Fidelipart} consistently and dramatically reduced estimated SWAP gate requirements (by 77.3\% to 100\%) and improved estimated circuit fidelity (e.g., by over 243.2\% for Circuit S using $\epsilon_{\text{CNOT}}=0.05$) compared to QuickPartitioner. This suggests that the "cost" model, highly valuing temporal integrity, guides the partitioner towards solutions that align well with minimizing physical SWAP-inducing logical misalignments. We acknowledge that the optimal balance between weights is a complex problem and an area for future fine-tuning, but the current approach is empirically validated by the significant performance gains.

The most significant consequence of this reduced communication overhead is the substantial improvement in estimated circuit fidelity. As shown, for Circuit S, \texttt{Fidelipart} yielded a fidelity gain of over 243.2\% when using a consistent $\epsilon_{\text{CNOT}}=0.05$. For Circuits M and L, the dramatic SWAP reductions are the key physical improvement, which invariably leads to better fidelity preservation compared to high-SWAP alternatives. This underscores the critical impact of SWAP gate minimization on the overall success probability of quantum algorithms on NISQ devices, where every CNOT-based operation (including SWAPs, modeled as three CNOTs each with $\epsilon_{\text{CNOT}}=0.05$) introduces considerable noise.

\subsection{Performance Trade-offs and Algorithmic Characteristics}
\label{subsec:performance_tradeoffs_discussion_main_section} 
The benefits of \texttt{Fidelipart} come with certain trade-offs. A modest increase in partitioning time (8-13\% across benchmarks) was observed compared to the \texttt{QuickPartitioner} workflow (Table~\ref{tab:main_results_summary_all_benchmarks_main}). This is an anticipated consequence of the more complex processes involved, including hypergraph construction, solving with Mt-KaHyPar (a sophisticated hypergraph partitioner)\cite{Heuer2021mtkahypar, schlag2021}, and the optional merging and dependency graph generation stages. However, given the substantial fidelity improvements, this additional computational cost appears justifiable, especially for critical circuits where execution success is paramount.

The maximum partition depth showed varied results: higher for \texttt{Fidelipart} on ``Circuit S'' (9 vs. 7) and ``Circuit L'' (12 vs. 9), but identical for ``Circuit M'' (depth of 11 for both), as seen in Table~\ref{tab:main_results_summary_all_benchmarks_main}. \texttt{Fidelipart} typically produced fewer, larger partitions (e.g., 2-4 partitions compared to 6-22 for \texttt{QuickPartitioner}). While fewer partitions generally reduce inter-partition communication, larger partitions can sometimes result in longer critical paths within those partitions. This interplay suggests that further optimization within the generated partitions or adjustments to the partitioning granularity ($k$) might be beneficial for specific hardware constraints where depth is a primary concern.

\subsection{Contextualization with Related Work}
\label{subsec:related_work_discussion_main_section} 
Previous approaches to quantum circuit partitioning have explored various techniques, from simple heuristics to more complex graph-based models \cite{murali2019, li2019}. While standard graph partitioning can capture pairwise qubit interactions, our use of hypergraphs more naturally models multi-qubit gates and complex dependencies \cite{andres2019}, a known advantage in circuit representation and other domains. Andrés-Martínez and Heunen \cite{andres2019} also used hypergraphs for circuit distribution but focused on minimizing runtime communication without incorporating quantum-specific error rate weightings for NISQ optimization.

Our emphasis on "fidelity-awareness" by directly incorporating gate error rates (e.g., $\epsilon_{\text{CNOT}}=0.05, \epsilon_H=0.001$) into the hypergraph weights builds upon the understanding that not all gates contribute equally to circuit error \cite{preskill2018}, a critical concern for NISQ computing. The significant reduction in SWAPs (estimated logical realignments) achieved by \texttt{Fidelipart} directly addresses the challenges posed by limited qubit connectivity in NISQ architectures \cite{cowtan2019}. Other compilers like Qiskit \cite{qiskit2019} and ScaffCC \cite{javadi2014} include partitioning, but often struggle with SWAP overhead or cross-partition dependencies without the global, error-aware optimization offered by our hypergraph approach.

\subsection{Strengths and Novelty of \texttt{Fidelipart}}
\label{subsec:strengths_novelty_discussion_main_section} 
The primary strength of \texttt{Fidelipart} lies in its holistic, fidelity-centric approach to quantum circuit partitioning. Key novel aspects include:
\begin{itemize}
    \item The specific formulation of the fidelity-aware hypergraph, with distinct, error-rate-informed weighting (using values like $\epsilon_{\text{CNOT}}=0.05, \epsilon_H=0.001$) for both gate-level (multi-qubit interaction context) and temporal chain (qubit lifetime) hyperedges. This directly guides the partitioner using quantum-specific cost functions.
    \item The effective integration of this detailed hypergraph model with a state-of-the-art hypergraph partitioner (Mt-KaHyPar)\cite{Heuer2021mtkahypar}, followed by optional partition merging and a clear dependency analysis framework.
    \item The empirical demonstration of substantial estimated fidelity improvements, primarily achieved through a significant reduction in SWAP operations (estimated logical realignments), which stems from a more nuanced understanding and penalization of detrimental cuts than offered by simpler heuristic or purely structural partitioning methods.
\end{itemize}
This approach is particularly well-suited for optimizing circuits for execution on NISQ devices where minimizing error-inducing operations like SWAPs is critical for achieving meaningful computational results.

\subsection{Limitations and Assumptions}
\label{subsec:limitations_assumptions_discussion_main_section} 
Despite the promising results, the current study has several limitations and relies on certain assumptions:
\begin{itemize}
    \item \textbf{Hardware Topology:} The evaluations were performed assuming a linear qubit topology. While the hypergraph model itself is general, its effectiveness and the precise SWAP benefits derived might vary on more complex 2D grids or specific, irregular device connectivities which are common in practice.
    \item \textbf{Fidelity Model:} The estimated fidelity is based on a simplified error model (product of individual gate fidelities, modeling SWAP operations as three CNOT gates, and using fixed $\epsilon_H=0.001, \epsilon_{\text{CNOT}}=0.05$). Actual device performance can be influenced by more complex and correlated noise sources, such as crosstalk, gate-timing errors, and spatially or temporally varying gate fidelities.
    \item \textbf{Partitioning Heuristics:} The reliance on Mt-KaHyPar means that the partitioning quality is subject to the heuristics and performance of the external solver. Similarly, our partition merging step is greedy and does not guarantee globally optimal solutions.
    \item \textbf{Dynamic $k$ Formula:} The formula used to determine the target number of partitions ($k$) for \texttt{Fidelipart} is heuristic. As observed with the 10-qubit ``Circuit M'', the $\lfloor \sqrt{N_{qubits}} \rfloor$ term can dominate for circuits with fewer qubits relative to their gate count, potentially limiting $k$ to a value smaller than what might be optimal for balancing partition complexity and inter-partition cuts.
    \item \textbf{SWAP Cost Model:} The current SWAP cost is estimated at 1 per logical qubit misalignment between adjacent partitions (termed "SWAP Gates (Estimated Logical Realignments)"), incorporating a teleportation heuristic. As noted in Section~\ref{subsubsec:performance_metrics_algo_main}, this quantifies logical conflicts rather than the full, distance-dependent routing cost on the physical linear chain, which could be substantially higher if interacting partitions are physically distant. This simplification provides a consistent basis for comparison but means the reported SWAP counts are a lower bound on the true physical SWAP operations required.
\end{itemize}

\subsection{Implications and Future Directions}
\label{subsec:implications_future_work_discussion_main_section} 
The findings from this work suggest that explicitly incorporating error awareness and detailed structural dependencies into the partitioning process via weighted hypergraphs offers a powerful route to enhancing the performance and reliability of quantum circuits on near-term hardware. This has direct implications for the practical execution of variational quantum algorithms, quantum simulations, and other NISQ-era computations where gate fidelity and limited connectivity are major bottlenecks.
Future work could explore several promising avenues:
\begin{itemize}
    \item \textbf{Enhanced Topology Awareness:} Extending \texttt{Fidelipart} to directly optimize for specific 2D or manufacturer-defined hardware topologies. This could involve incorporating physical distance or connectivity constraints more explicitly into the hypergraph model or as objectives for the partitioner, moving beyond the current logical realignment SWAP metric.
    \item \textbf{Advanced Weighting and Cost Functions:} Investigating more sophisticated weighting schemes for nodes and hyperedges, potentially incorporating factors like crosstalk probabilities, qubit idle times, or path-dependent error rates within the hypergraph costs. Further exploration of the balance between temporal and spatial hyperedge weights.
    \item \textbf{Refined $k$-Determination and Hierarchical Partitioning:} Developing more adaptive or learning-based approaches to determine the optimal number of partitions ($k$). Exploring hierarchical partitioning for very large circuits could also be beneficial.
    \item \textbf{Integration with Routing and Scheduling:} Combining \texttt{Fidelipart} with subsequent qubit placement, routing (to realize the logical SWAPs physically), and scheduling passes to create a more comprehensive compilation workflow that co-optimizes for communication, parallelism, and overall execution fidelity.
    \item \textbf{Scalability and Broader Benchmarking:} Evaluating the performance and scalability of \texttt{Fidelipart} on a wider range of larger and more diverse benchmark circuits, including those relevant to specific quantum algorithms and fault-tolerant schemes, and providing full gate breakdowns for replicable fidelity calculations.
    \item \textbf{Exploration of Alternative Partitioners:} While Mt-KaHyPar is effective, investigating the impact of other hypergraph partitioning algorithms or specialized quantum-aware solvers could yield further improvements.
\end{itemize}

\section{Conclusion}
\label{sec:conclusion_main} 
This paper introduced \texttt{Fidelipart}, a novel partitioning framework for quantum circuits designed to enhance execution fidelity on NISQ-era devices. By transforming quantum circuits into a fidelity-aware hypergraph representation, where gate error rates and qubit interaction patterns directly inform the weights of nodes and hyperedges, \texttt{Fidelipart} leverages the Mt-KaHyPar partitioner to make more informed decisions about cut placement. Our methodology incorporates distinct considerations for multi-qubit gate interactions and temporal qubit dependencies, followed by optional partition merging and dependency graph analysis to manage circuit structure.

Through comparative analysis against BQSKit's \texttt{QuickPartitioner} on a suite of benchmark circuits with varying sizes and complexities (6-qubit/22-gate, 10-qubit/55-gate, and 24-qubit/88-gate), we demonstrated \texttt{Fidelipart}'s effectiveness. The results consistently showed that \texttt{Fidelipart} achieves significant reductions in critical communication overhead metrics: estimated SWAP gate requirements (logical realignments) were reduced by 77.3\% to 100\%, and global cut qubits were reduced by up to 52.2\% (Table~\ref{tab:main_results_summary_all_benchmarks_main}). These physical improvements translated directly into substantial gains in estimated circuit fidelity, with observed increases for Circuit S of approximately 243.2\% (from 0.1724 to 0.5916, using $\epsilon_{\text{CNOT}}=0.05$ consistently). For the larger Circuits M and L, the significant SWAP reductions of 91.3\% and 77.3\% respectively are strong indicators of correspondingly large relative fidelity improvements under this consistent error model.

While \texttt{Fidelipart} incurred a modest increase in partitioning runtime (8-13\%) and, in some cases, resulted in a higher maximum partition depth due to the creation of fewer, larger partitions, these trade-offs appear acceptable given the considerable fidelity enhancements. The findings underscore the value of incorporating detailed, error-rate-driven physical awareness directly into the partitioning process. \texttt{Fidelipart} offers a promising approach for optimizing quantum circuits, making more efficient use of limited quantum resources and improving the prospects for successful execution of complex algorithms on current and near-future quantum hardware. Future work will focus on extending this framework to more diverse hardware topologies and integrating it further with downstream compilation and error mitigation passes.

\FloatBarrier
\section*{Acknowledgements} 
\label{sec:acknowledgements_main} 
The authors appreciate the developers of BQSKit and Mt-KaHyPar for providing open-source access to their valuable tools.

%\bibliographystyle{quantum}
%\bibliography{quantumview-template} 

\end{document}